\documentclass[pre,twocolumn,groupedaddress,showpacs,showkeys,amsmath,amssymb,floatfix]{revtex4}
\usepackage{graphicx}
\usepackage{dcolumn}
\usepackage{bm}
\begin{document}

\title{Fragmentation properties of two-dimensional Proximity Graphs considering random failures and targeted attacks}
\author{C. Norrenbrock}
\email{christoph.norrenbrock@uni-oldenburg.de}
\author{O. Melchert}
\email{oliver.melchert@uni-oldenburg.de}
\author{A. K. Hartmann}
\email{alexander.hartmann@uni-oldenburg.de}
\affiliation{Institut f\"ur Physik, Universit\"at Oldenburg, 26111 Oldenburg, Germany}
\date{\today}

\begin{abstract}

The pivotal quality of proximity graphs is connectivity, i.e.\ all nodes in the
graph are connected to one another either directly or via intermediate nodes.
These types of graphs are robust, i.e., they are able to function well even if
they are subject to \emph{limited} removal of elementary  building blocks, as
it may occur for  random failures or targeted attacks.  Here, we study how the
structure of these graphs is affected when nodes get removed successively until
an extensive fraction is removed such that the graphs fragment.  We study
different types of proximity graphs for various node removal strategies.  We
use different types of observables to monitor the fragmentation
process, simple ones like number and sizes of connected components, and more
complex ones like the hop diameter and the backup capacity, which
is needed to make a network $N-1$ resilient.  The actual fragmentation
turns out to be described by a second order phase transition.  Using
finite-size scaling analyses we numerically assess the threshold fraction of
removed nodes, which is characteristic for the particular graph type and node
deletion scheme, that suffices to decompose the underlying graphs. 

\end{abstract}

\maketitle

\section{Introduction \label{sect:introduction}}

The pivotal issue of standard percolation \cite{stauffer1979,Stauffer1992} is
connectivity. A basic example is $2D$ random site percolation, where
one studies a lattice in which a random fraction of the sites is
``occupied''.  Clusters composed of adjacent occupied sites are then analyzed
regarding their geometric properties.  Depending on the fraction $p$ of
occupied sites, the geometric properties of the clusters change, leading from a
``fragmented'' phase with rather small and 
disconnected clusters to a phase, where there is basically one large 
connected cluster covering the lattice.  Therein, the appearance of
an infinite, i.e.\ percolating, cluster is described by a second-order phase
transition.  

Similar to the issue of connectivity is the robustness, i.e.\ 
the ability of networks 
to function well even if they are subject to random failures or targeted
attacks of their elementary building blocks, e.g., node removal. 
This is of particular importance for more applied 
real-world networks, which may fail even if they are still connected, e.g.,
when the dynamics of nodes is not synchronous due to an failure.
Also, many real-world networks are not  embedded in two-dimensions,
they may even exhibit an infinite-dimensional, i.e., mean-field structure.
E.g., electrical power grids must ensure power supply for entire 
resident population \cite{Albert2004}, 
urban road networks \cite{Jiang2004} and airline networks \cite{Choi2006} facilitate social 
and economical interaction, and the internet \cite{Barabasi2000}, which has become essential in almost all aspects of life.
In general, networks are represented by a set of nodes, i.e.\ the elementary
building blocks of a network, and pairs of nodes might be joined by edges. One
possibility to characterize a network (or graph for that matter) is by means of
its degree distribution, where the degree of a node refers to the number of its
adjacent neighbors.  Several real-world networks, such as the hyperlink-network
of the internet, exhibit a scale-free degree distribution \cite{Barabasi2000}.
During the last decade, various studies have been published that focus on this
prototypical type of degree distribution.  In particular, the fragmentation
properties of scale-free Barab\'{a}si-Albert (BA) networks \cite{Albert2000,Callaway2000,Crucitti2004-1,Gallos2005,Holme2002,Cohen2001} (and also of several other ones \cite{Huang2011,Kurant2007,Paul2004,Shargel2003,Tanizawa2012}) have been put under
scrutiny.
In the aforementioned articles, different node-removal strategies have been
considered to investigate the fragmentation properties of the considered
networks.  It turns out that scale-free networks are robust against random node
removals, but very vulnerable to intentional attacks targeting particular
``important'' nodes.  Note that there are many different local and global
measures to quantify whether a node is important. Popular choices are, e.g., the
degree of a node, its betweenness-centrality \cite{Brandes2008} (subject to a
particular metric used to measure the length of shortest paths between pairs of
nodes), or, somewhat more specific to the hyperlink structure of the internet,
the ``PageRank'' \cite{Page1999} relevance measure for web pages.

In the presented work we focus on types of networks, which 
are completely different from scale free graphs. The networks considered here 
are constructed from sets of points distributed in the two-dimensional 
Euclidean
plane. More precisely, we consider three types of \emph{proximity graphs},
namely \emph{relative neighborhood graphs} (RNGs) \cite{Toussaint1980},
\emph{Gabriel graphs} (GGs) \cite{Gabriel1969}, and \emph{Delaunay
triangulations} (DTs) \cite{Sibson1978}.  These are planar graphs
\cite{Essam1970} where pairs of nodes are connected by undirected edges if they
are considered to be close in some sense (see definitions in Sec.\
\ref{sect:proxigraph}). In addition, we consider also a certain type of (non
planar) geometric random network, termed \emph{minimum-radius graph} (MR),
where pairs of nodes are connected if their distance does not exceed a
particular threshold value.
The above proximity graphs where already studied in different scientific fields
such as the simulation of epidemics \cite{Toroczkai2007}, percolation
\cite{Bertin2002,Becker2009,Billiot2010,melchert2013}, and message routing and
information dissemination in ad-hoc networking
\cite{Jennings2002,Santi2005,Li2005,Rajaraman2002}.  To elaborate on the latter
point, proximity graphs find application in the construction of planar
``virtual backbones'' for ad-hoc networks, i.e.\ collections of radio devices
without fixed underlying infrastructure, along which information can be
efficiently transmitted \cite{Karp2000,Bose2001,Jennings2002,Yi2010,Kuhn2003}.
Routing with guaranteed node-to-node connectivity (at least in a multi-hop
manner) is especially important to ensure a complete broadcast of information
in ad-hoc networks \cite{Jennings2002}.  Here, we consider three types of node
removal strategies with different levels of severity, see Sec.\
\ref{sect:strat}, and we numerically assess the threshold fraction of removed
nodes (characteristic for the particular graph type and node deletion scheme)
that suffices to decompose the underlying graphs into ``small'' clusters. 

The remaining article is organized as follows.  In Sec.\ \ref{sect:proxigraph}
we introduce the four different graph types that were considered in the
presented study.  In Sec.\ \ref{sect:strat} we describe the three node-removal
strategies that were used in order to characterize the fragmentation process
for each of these graph types.  In Sec.\ \ref{sect:results} we introduce the
observables that were recorded during the fragmentation procedure and we list
the results of our numerical simulations. Finally, Sec.\ \ref{sect:conclusion}
concludes with a summary.

\begin{figure}[t]
\centerline{
\includegraphics[width=0.85\linewidth]{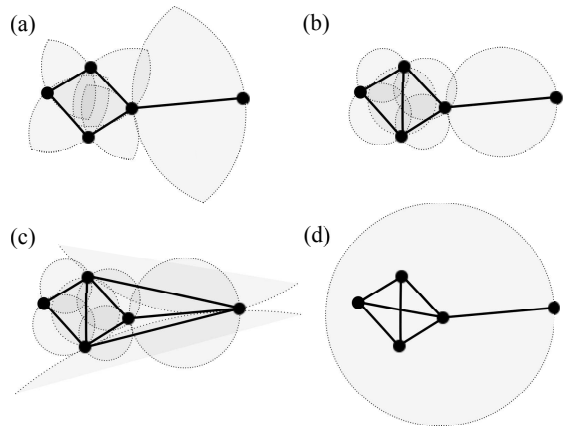}}
\caption{
Examples of the four different graph types for a small set of $N=5$ nodes (see text for details).
(a) instance of a RNG, where for all pairs of nodes that will be connected under the respective linking-rule, the respective lune is depicted in gray.
(b) instance of a GG, where for all pairs of nodes that will be connected under the respective linking-rule, the circle that helps in the decision making process is depicted in gray. 
(c) instance of a DT, where the gray shaded circles are exemplary for those that might aid in the decision making process. 
(d) instance of a MR, where the linking range $r$ is depicted (gray circle) for a single node only (all other nodes exhibit the same linking range).
\label{fig:sketches}}
\end{figure}

\section{Graph Types \label{sect:proxigraph}}

Subsequently we introduce four different types of graphs for a planar set of,
say, $N$ points and we characterize the fragmentation process on each of these
graph types following three different node-removal strategies, detailed in
Sec.\ \ref{sect:strat}.  Three of these graph types, introduced in Subsects.\
\ref{subsect:RNG} through \ref{subsect:DT} belong to the class of
\emph{proximity graphs} \cite{Jaromczyk1992}. The fourth graph type, detailed
in Subsect.\ \ref{subsect:MR}, is a particular type of a \emph{random geometric
graph}.  Below, a graph is referred to as $G=(V,E)$, where $V$ comprises its
node-set ($N=|V|$; where $N$ is also referred to as ``system size''), and where
$E$ ($M=|E|$) signifies the respective edge-set \cite{Essam1970}.  Each of the $N$ nodes
$u\in V$ represents a point in the two-dimensional unit square for which the
coordinates $u_x$ and $u_y$ are drawn uniformly and independently
at random.  So as to compute
the distance ${\rm dist}(u,v)$ between two nodes $u,v \in V$ we consider the
Euclidean metric under which ${\rm dist}(u,v)=[(u_x-v_x)^2+(u_y-v_y)^2]^{1/2}$.
We further consider open boundary conditions. Thus an increase of the system size 
corresponds to increasing the density of nodes on the unit square. On the
other hand, so as to maintain the density of nodes while increasing $N$, the
networks can be pictured as having an effective side-length $L=\sqrt{N}$. A
common feature of these four types of graphs is that their edge-set encodes
proximity information regarding the close neighbors of the terminal nodes of a
given edge.  The different graph types can be distinguished by the precise
linking-rule that is used to construct the edge-set for a given set of nodes. In
this section the linking-rules that define the four types of proximity graphs
will be detailed.

\subsection{Relative Neighborhood Graphs (RNGs)}\label{subsect:RNG}

One particular proximity graph type that will be considered subsequently is the
\emph{relative neighborhood graph} (RNG) \cite{Toussaint1980}.  In order to
determine whether in the construction procedure for an instance of a RNG two
nodes $u,v\in V$ need to be connected to each other, it is necessary to check
if there is a third node $w\in V\setminus\{u,v\}$ with ${\rm dist}(u,w) \leq
{\rm dist}(v,u)$ and ${\rm dist}(v,w) \leq {\rm dist}(v,u)$.  If such a node
$w$ does not exist, $u$ and $v$ will get linked.  In geometrical terms, for
each pair $u$ and $v$ of points, the respective distance $\rm{dist}(u,v)$ can
be used to construct the \emph{lune} ${\rm lune}(u,v)$.  The lune is given by
the intersection of two circles with equal radius $\rm{dist}(u,v)$, centered at
$u$ and $v$, respectively.  If no other point $w \in V\setminus\{u,v\}$ lies
within ${\rm lune}(u,v)$, i.e.\ if the lune is empty, both nodes are connected
by means of an edge.  To facilitate intuition, an example of a RNG for a small
set of $N=5$ nodes is sketched in Fig.\ \ref{fig:sketches}(a).  A larger
example that illustrates the principal structure of a RNG is shown in Fig.\
\ref{fig:topo}(a).

\subsection{Gabriel Graphs (GGs)}\label{subsect:GG}

Another proximity graph that will be considered in this article is the
\textit{Gabriel graph} (GG) \cite{Gabriel1969,Bertin2002}.  To determine
whether in the construction procedure for an instance of a GG two nodes $u,v
\in V$ need to be connected, $\rm{circ}(u,v)$, i.e.\ the smallest of all
possible circles which embeds both nodes is considered, which has a diameter
$\rm{dist}(u,v)$.  These two nodes will be connected unless there is another
node $w$ which is located within the area enclosed by ${\rm circ}(u,v)$.  To
facilitate intuition, the linking rule for the GG is illustrated in Fig.\
\ref{fig:sketches}(b).  A larger example that illustrates the principal
structure of a GG is shown in Fig.\ \ref{fig:topo}(b).  Further, note that the
GG is a super-graph of the RNG.  This is due to the circumstance that
$\rm{circ}(u,v)$, which is relevant in the construction procedure of a GG
instance for a given set of nodes encloses a subarea of $\rm{lune}(u,v)$, being
relevant in the construction procedure of the corresponding RNG instance
(compare the grey shaded surfaces in Figs.\ \ref{fig:sketches}(a,b)).
Therefore, all edges contained in the RNG are also included in the GG.  Note
that this can also be seen in Figs.\ \ref{fig:topo}(a,b).

\subsection{Delaunay Triangulations (DTs)}\label{subsect:DT}

\begin{figure}[t]
\centerline{
\includegraphics[width=0.9\linewidth]{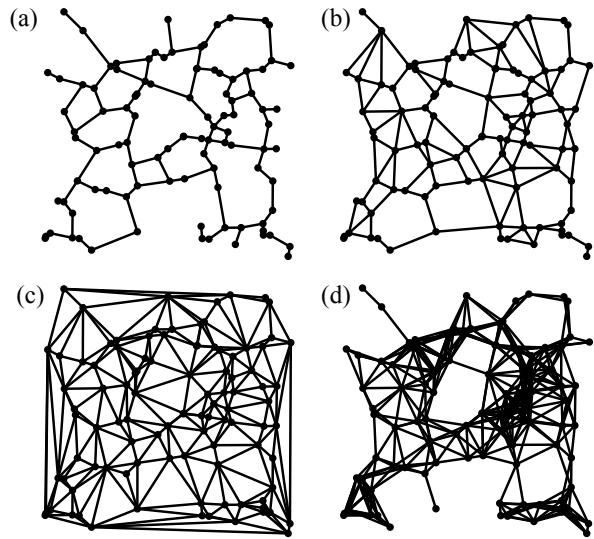}}
\caption{    
Examples of the four different graph types for the same set of $N=100$ nodes, distributed uniformly at random in the 2D unit square.
(a) RNG, 
(b) GG,
(c) DT, and
(d) MR.
\label{fig:topo}}
\end{figure}

The construction of the \textit{Delaunay triangulation} (DT; also a type of
proximity graph) \cite{Sibson1978} is quite similar.  Two nodes $u,v \in V$
will be connected if any circle exists which embeds $u$ as well as $v$ but no
further nodes.  To facilitate intuition, the result of this linking-rule is
shown in Fig.\ \ref{fig:sketches}(c).  A larger example that illustrates the
principal structure of a DT is shown in Fig.\ \ref{fig:topo}(c).  From the
definition of these linking-rules, since the GG graph also involves the
construction of a circle, it is evident that an instance of a DT for a given
set of nodes must be a super-graph of the corresponding GG instance.  As a
consequence, being a sub-graph of the GG, the RNG is also a sub-graph of the
DT.  This can be observed in Figs.\ \ref{fig:sketches}(a-c) (Figs.\
\ref{fig:topo}(a-c)), where the RNG, GG and DT are illustrated for the same set
of $N=5$ ($100$) nodes.

\subsection{Minimum Radius Graphs (MRs)}\label{subsect:MR}

The fourth network topology that will be considered is the \textit{minimum
radius graph} (MR). In the construction procedure of an instance of a MR, two
nodes $u,v \in V$ will be joined by an edge, if ${\rm dist}(u,v)\leq r$.
Therein, the ``connectivity radius'' $r$ specifies 
the smallest possible value which ensures
that all nodes are connected to one another, possibly in a multi-hop manner.
It becomes evident from Figs.\ \ref{fig:sketches}(d) and \ref{fig:topo}(d)
that, in contrast to the previous graphs, the MR might feature crossing edges.

\subsection{Graph construction}\label{subsect:algs}

In order to construct the RNG and GG, we made use of the sub-graph hierarchy
${\rm RNG} \subset {\rm GG} \subset {\rm DT}$. I.e., for a given set of nodes
we first obtained the DT by means of the {\rm Qhull} computational geometry
library \cite{qhull} (the DT for a set of $N$ points can be computed in time
$O(N\log(N))$ \cite{CompGeom1985,qhull}) and then pruned the resulting edge-set
$E$ until the linking requirements of GG or RNG are met.  Here, we amend the
naive implementation of this two-step procedure \cite{Toussaint1980}, yielding
an algorithm with running time $O(N^2)$, by means of the ``cell-list'' method
\cite{melchert2013}, resulting in a sub-quadratic running time.  In this
regard, note that Ref.\ \cite{Jaromczyk1992} provides an overview of several
algorithmic approaches for the construction of RNGs and GGs.  
Finally, note that RNGs and GGs can be found as the limiting cases of a
parameters family of proximity graphs, termed $\beta$-skeletons \cite{Kirkpatrick1985}.

At this point, note that due to a yet unmentioned property of minimum weight
spanning trees (MST; i.e.\ a spanning tree in which the sum of Euclidean edge
lengths is minimal, see Ref.\ \cite{Cormen2001}) we can set the ``connectivity
radius'' of MRs, i.e.\ a geometric random graph, in context to proximity
graphs.  Bear in mind that the longest edge present in any instance of a MR
specifies the smallest possible edge length which ensures that all nodes are
connected to one another. Exactly this edge length characterizes the longest
edge in the MST of the corresponding node-set. For a given set of nodes, a MST
is a spanning sub-graph of the RNG \cite{Toussaint1980,melchert2013}.  Thus,
considering MSTs, the previously mentioned sub-graph hierarchy can be extended
to ${\rm MST} \subset {\rm RNG} \subset {\rm GG} \subset {\rm DT}$. This allows
for a fast construction of a MR instance for a given set of points via a
convenient three-step procedure: (i) compute the DT for the given set of
points, (ii) filter the edge-set of the DT to determine the corresponding MST,
and, (iii) use the length of the longest MST edge as ``connectivity radius'' to
construct the respective MR. Therein, the overall running time is dominated by
step (iii), which, in its most naive implementation has computational cost
$O(N^2)$.  Note that during the latter step, the previously mentioned
``cell-list'' method can be used to achieve an improved running time.

Subsequently we will introduce the node-removal strategies that will be considered
in the numerical simulations carried out to characterize the fragmentation process
for the above graph types.

\section{Node-removal strategies\label{sect:strat}}

As pointed out above, in the presented article we aim at characterizing the
fragmentation processes for the graph types introduced in Sec.\
\ref{sect:proxigraph}. Therefore we consider three different types of
node-removal strategies that are used throughout the literature
\cite{Albert2000,Callaway2000,Crucitti2004-1,Gallos2005,Holme2002}. For
convenience these will be detailed subsequently.  Therefore, note that the
basic procedure to study the fragmentation process for a single network
instance consists in successively removing nodes until the network is
decomposed into many small clusters of nodes, thereby recording observables
that provide information about the current characteristics of the network (see
Sec.\ \ref{sect:results}). 

The most simplistic node-removal strategy followed here is termed \emph{random
failure}. According to this strategy, a node is picked uniformly at random and
deleted from the network (along with all its incident edges).

Depending on the context into which the networks are set, it might be useful to
associate a measure of \emph{relevance} to each node. Then it is also intuitive
to ask for node-removal strategies that preferentially target the most relevant
nodes. Removal strategies that capitalize on the relevance of a node are termed
\emph{targeted attacks}. Here, we consider two different targeted attack
strategies

(i)
degree-based attack (conveniently abbreviated as ``attack 1''), where the relevance of a node is simply
measured by its degree (i.e.\ the number of its incident edges). The higher the
degree of a node, the more relevant it is assumed to be.  Accordingly, at each
elementary node removal step during the fragmentation process, the node with
the currently highest degree is selected for deletion.  If, at a given step,
there are many nodes exhibiting the currently highest degree, one of these
nodes is chosen uniformly at random. Note that the degree of a node is a local
property only, i.e.\ for a given node one only has to determine the number of
its nearest neighbors. Thus, from a computational point of view the node degree
is a very inexpensive relevance measure.  

(ii) betweenness-based attack 
(conveniently abbreviated as ``attack 2''), where
the relevance of a node is measured by its betweenness centrality
\cite{Brandes2008}.  The betweenness centrality of node $u$ is the number of
shortest paths between all node pairs $(v,w)$ $(v,w\neq u)$ that pass through
$u$.  The larger the value of the betweenness centrality, the more relevant a
node is assumed to be.  In some applications, the Euclidean distance along the
edges is relevant for determining shortest paths \cite{Cormen2001}.  However,
here we instead considered the hop-metric, where distances are simply measured
in terms of node-to-node hops. Consequently, the shortest path problem can be
solved by means of a breadth-first search \cite{Cormen2001}.  During each
elementary node-removal step, 
the node exhibiting the currently highest value of
betweenness centrality gets removed. As before, if several nodes have the same
value, one of them is chosen uniformly at random.  Note that the
betweenness centrality is a global property deduced from the underlying
network, i.e.\ for the betweenness centrality of a particular node, the
configuration of shortest paths between all pairs of nodes is of relevance.
From a computational point of view this is, of course, considerably more
expensive than the computation of the local node degree.

Subsequently, we will use the above node-removal strategies in order to
characterize the fragmentation process for the graph types described in Sec.\
\ref{sect:proxigraph} by means of numerical simulations.

\section{Results\label{sect:results}}

In the current section we will report on numerical simulations for the
different graph types for planar sets of $N=144 (=12^2)$ up to $36864(=192^2)$
points, where results are averaged over 2000 independent graph instances.  In
Sec.\ \ref{subsect:topology} we first report on some topological properties of
the graphs, in Sec.\ \ref{subsect:fragmentationProcedure} the analysis of the
fragmentation procedure is summarized. In Sec.\ \ref{subsect:stability} further
issues concerning the resilience of the networks seen as transport networks
(``$N-1$ stability'') are discussed. Finally, in Sec.\ \ref{subsect:stability},
the networks will be compared under the assumption that they all
exhibit the same summed-up edge length.

Subsequently, albeit we will present results for all relevant combinations of
the four graph types and three node-removal strategies, we will not show
figures with results for all these combinations.  Instead, so as to illustrate
the analyses performed in the following section, we mainly present figures for
the RNG proximity graphs subject to a degree-based node removal strategy. 

\subsection{Topological properties}\label{subsect:topology}

\begin{figure}[t!]
\centerline{
\includegraphics[width=1.0\linewidth]{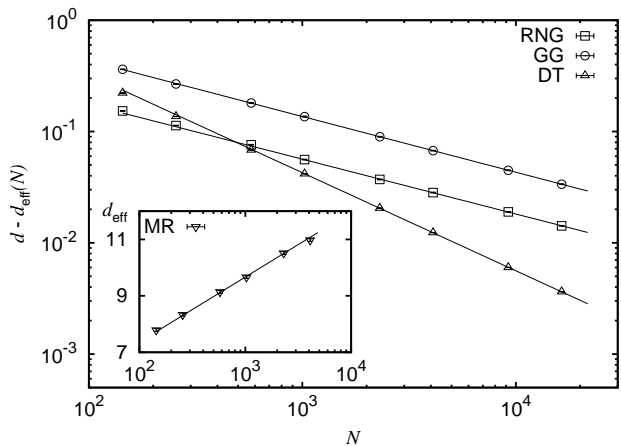}}
\caption{Finite-size scaling behavior of the average degree for the four
different graph types. The main plot shows the scaling behavior 
found for the three types of proximity graph, i.e.\ RNG, GG, and DT (see text for details).
The inset shows the logarithmic scaling found in case of the MR graphs (see text for details).
\label{fig:degree}} 
\end{figure}

To emphasize structural differences between the graph types of the sub-graph
hierarchy ${\rm RNG} \subset {\rm GG} \subset {\rm DT}$ we first consider the
respective average node degree.  Therefore, the scaling behavior of the
effective, i.e. system-size dependent, average degree $d_{\rm eff}(N)$ is
considered and analyzed using a fit to the function $d_{\rm eff}(N)=d-aN^{-b/2}$.
For the three graph types ${\rm RNG}$, ${\rm GG}$ and ${\rm DT}$ the fits yield
asymptotic degrees $d$ and scaling exponents $b$, where $d_{\rm RNG}=2.557(1)$
and $b_{\rm RNG}=0.99(4)$ (with a reduced chi-square $\chi^2_{\rm red}=0.87$;
note that both, the asymptotic average degree and the scaling exponent compare
well to the estimates reported in Ref.\ \cite{melchert2013}), $d_{\rm
GG}=3.999(1)$ and $b_{\rm GG}=1.00(1)$ (reduced chi-square $\chi^2_{\rm
red}=0.70$), $d_{\rm DT}=6.0001(1)$ and $b_{\rm DT}=1.76(1)$ (for a reduced
chi-square $\chi^2_{\rm red}=1.87$; note that the average degree of the $\rm DT$
is known to be $d_{\rm DT}=6$). In Fig.\ \ref{fig:degree} the correction to 
scaling, i.e.\ $d-d_{\rm eff}(N)\propto N^{-b/2}$, is shown for the three types
of proximity graphs. It is interesting to note that RNG and GG
exhibit a similar scaling, involving a correction of the form $N^{-1/2}$, whereas
the scaling behavior for the average degree for the DT graphs is governed by 
a significantly larger exponent.
Also, note that instances of the three types of proximity graphs are planar, i.e.\
there are no crossing edges. While the bounding cycles of the finite faces for
the instances of RNGs and GGs might consist of an even or odd number of edges, all 
inner faces for instances of DTs are bounded by three edges.

Further, for the minimum radius graph we found that the effective, average
degree fits best to a logarithmic scaling function of the form $d_{\rm
eff}(N)=\log(a N)$, see inset of Fig.\ \ref{fig:degree} where $a=15.7(4)$
(reduced chi-square $\chi^2_{\rm red}=1.42$; however, note that the data can
also be fit by a scaling function with a small power-law correction as above,
where $d_{\rm MR}\approx 57$ and $b_{\rm MR}\approx 0.04$). 

Regarding MRs, consider that the longest edge present in any instance of a MST
(which specifies the length of the longest edge in  the respective MR instance;
see discussion above) can by no means exceed the length of the longest edge of
any of its super-graphs. Due to the geometric restrictions imposed by going from
an instance of a DT to a RNG, it is thus plausible that the maximal edge length
found for any MR instance is much shorter than, say, for the corresponding DT
instance. This holds in particular for the case of open boundary conditions,
where the outer faces of the DT instances feature rather long edges, see Fig.\
\ref{fig:topo}(c).  For a set of 500 instances of point sets consisting of
$N=16384$ nodes (i.e.\ for systems of effective side length $L=128$) we found
that the longest edge length ratio $r_{\rm max}/L$ for the four graph types read
$r_{\rm max}^{\rm DT}/L=0.714(6)$, 
$r_{\rm max}^{\rm GG}/L=0.02927(8)$,
$r_{\rm max}^{\rm RNG}/L=0.02335(8)$, and,
$r_{\rm max}^{\rm MR}/L=0.01548(6)$. 
For the first three graph types, these values should be more or less
independent of the system size. On the other hand,
 for the minimum radius graphs we found that the finite-size scaling
behavior of the connectivity radius $r_{\rm max}^{\rm MR}(L)$ as function of the effective system length $L$
exhibits a logarithmic scaling of the form $r_{\rm max}^{\rm
MR}(L)=a+b\log(L)$, where $a=1.334(8)$ and $b=0.133(2)$ (reduced chi-square
$\chi^2_{\rm red}=0.59$), supporting the logarithmic scaling of the average
degree. I.e., the respective ``connectivity area'' $A_{r} = \pi r_{\rm max}^2$,
which, if centered at the position of a given node, specifies the area in which
all its nearest neighbors can be found, should be almost
equal to the
previously discussed average degree $d_{\rm eff}$, because
the density of nodes is unity. E.g., at $N=2304$ (i.e.\
$L=48$) we find $A_r=10.7(2)$ and $d_{\rm eff}=10.52(5)$.

\begin{table}[b!]
\caption{\label{tab:critpoints}
Critical points $p_c$, i.e.\ fractions of removed nodes which indicate when the
underlying network decomposes into ``small'' clusters for the different graph
types and node removal strategies, discussed in Secs.\ \ref{sect:proxigraph}
and \ref{sect:strat} (random failure: equivalent to random percolation; attack
1: degree-based node removal strategy; attack 2: centrality-based node removal
strategy), respectively. 
} 
\begin{ruledtabular}
\begin{tabular}[c]{l@{\quad}llllll}
  strategy & RNG  & GG & DT & MR \\
  \hline
  random failure      & 0.205(1)   & 0.365(1)  & 0.500(2) & 0.71(1) \\
  attack 1 & 0.120(1)   & 0.263(1) & 0.377(1) & 0.68(2) \\
  attack 2 & 0  &  0 & 0 & 0 \\
  \end{tabular}
\end{ruledtabular}
\end{table}

\subsection{Fragmentation analysis}\label{subsect:fragmentationProcedure}

For the fragmentation analysis we consider instances of the four different
graph types, introduced in Sec.\ \ref{sect:proxigraph}, and successively
remove nodes according to one of the node removal strategies, presented in
Sec.\ \ref{sect:strat}, until the initially connected graph decomposes into
small clusters.  So as to determine the critical fraction $p$ of nodes that need to
be removed until the graph decomposes we perform a finite-size scaling (FSS)
analysis for different observables that are commonly used in studies of
percolation \cite{Stauffer1992} in Sec.\ \ref{subsect:percObs}. In
addition, in Sec.\ \ref{subsect:hopDiam} we consider the scaling behavior
of the hop-diameter, i.e.\ the longest among all shortest paths measured in
terms of node-to-node hops, which, e.g., is relevant in the context of
broadcasting problems on networks \cite{Jennings2002}.  

\subsubsection{Analysis of typical percolation observables}\label{subsect:percObs}

The observables we consider below can be rescaled following a common scaling
assumption.  Below, this is formulated for a general observable $y(p,N)$.  This
scaling assumption states that if the observable obeys scaling, it might be
written as
\begin{eqnarray}
y(p,L)= L^{-b}~f[(p-p_c) N^{1/(2\nu)}], \label{eq:scalingAssumption}
\end{eqnarray}
wherein $\nu$ and $b$ represent dimensionless critical exponents (or ratios
thereof, see below), $p_c$ signifies the critical point, and $f[\cdot]$ denotes
an unknown scaling function \cite{Stauffer1992,Binder2002}.   Following Eq.\
\ref{eq:scalingAssumption}, data curves of the observable $y(p,N)$ recorded at
different values of $p$ and $N$ \emph{collapse}, i.e.\ fall on top of each
other, if $y(p,N) N^{b/2}$ is plotted against $\epsilon \equiv (p-p_c) N^{1/(2\nu)}$
and if further the scaling parameters $p_c$, $\nu$ and $b$ that enter Eq.\
\ref{eq:scalingAssumption} are chosen properly.    The values of the scaling
parameters that yield the best data collapse determine the numerical values of
the critical exponents that govern the scaling behavior of the underlying
observable $y(p,N)$.  In order to obtain a data collapse for a given set of
data curves we here perform a computer assisted scaling analysis, see Refs.\
\cite{houdayer2004,autoScale2009}.

\begin{figure}[t!]
\begin{center}
\includegraphics[width=0.95\linewidth]{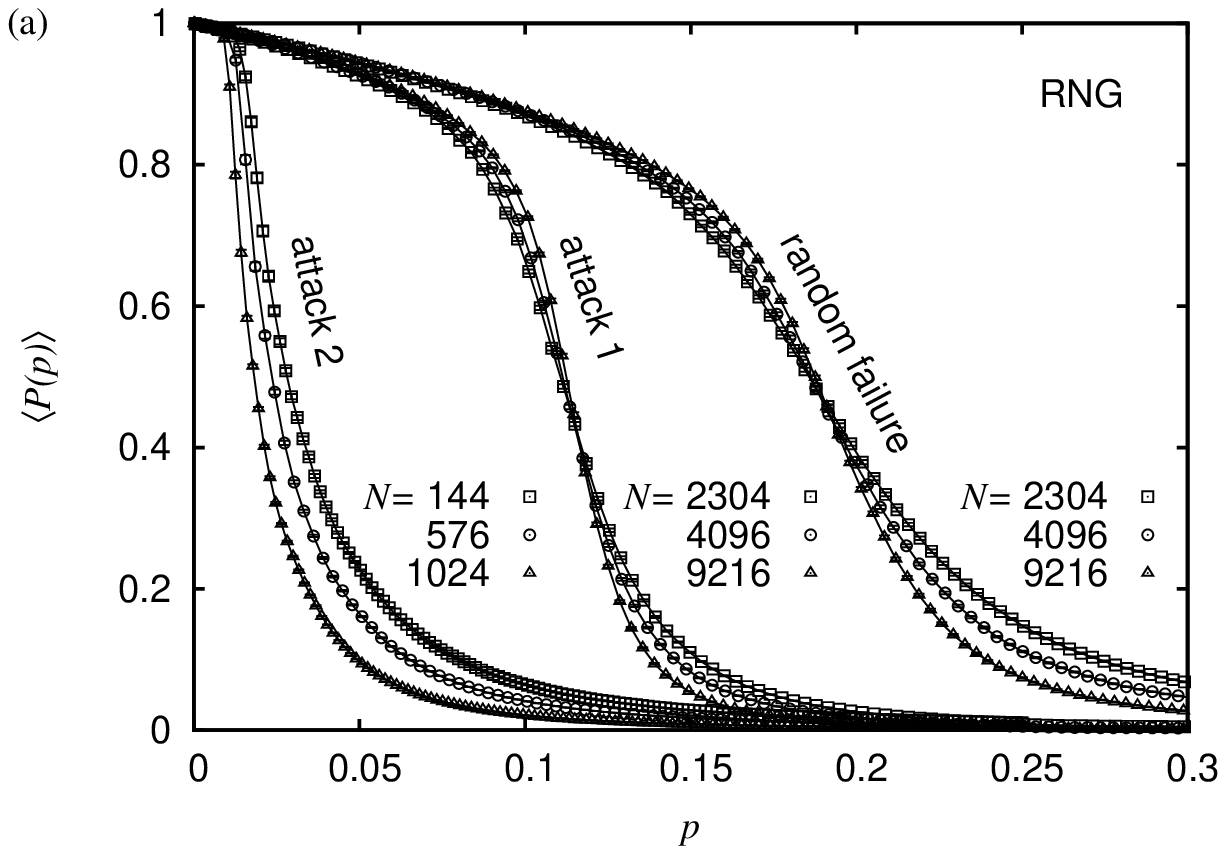}
\includegraphics[width=0.95\linewidth]{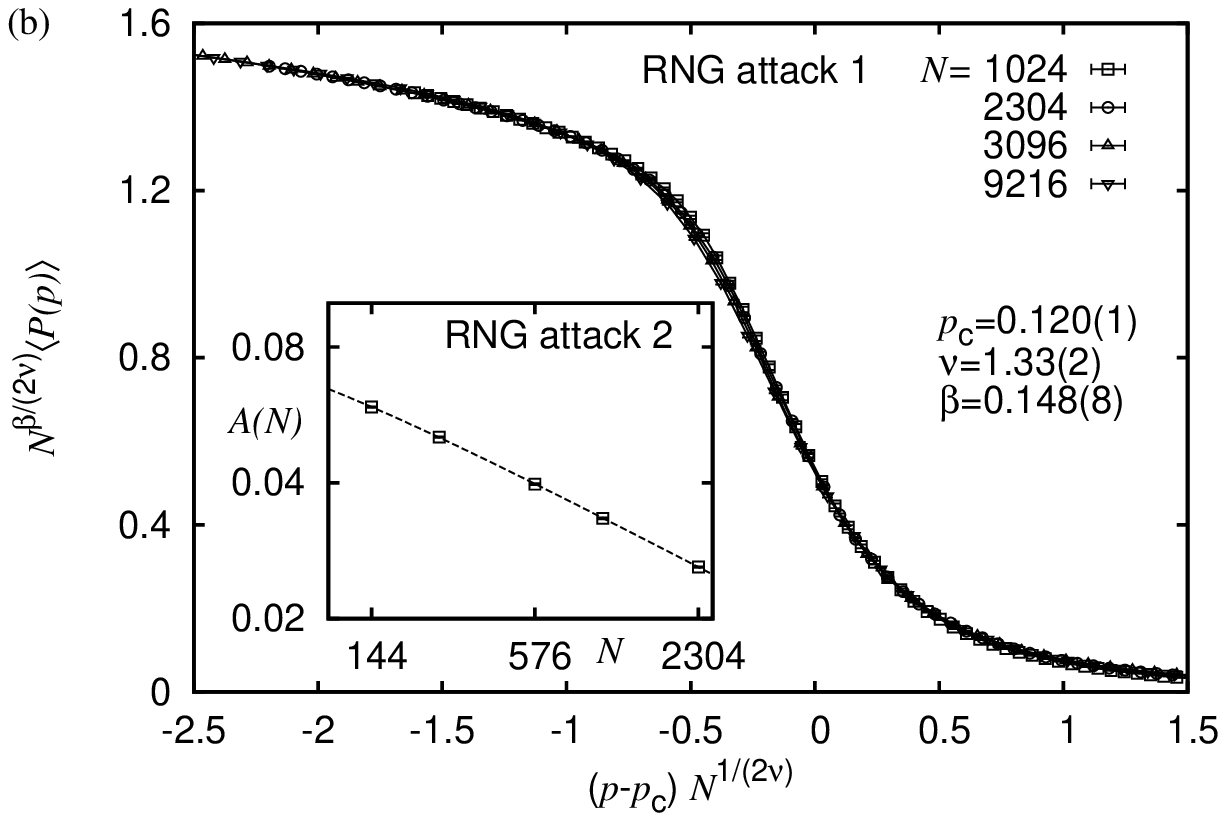}
\includegraphics[width=0.95\linewidth]{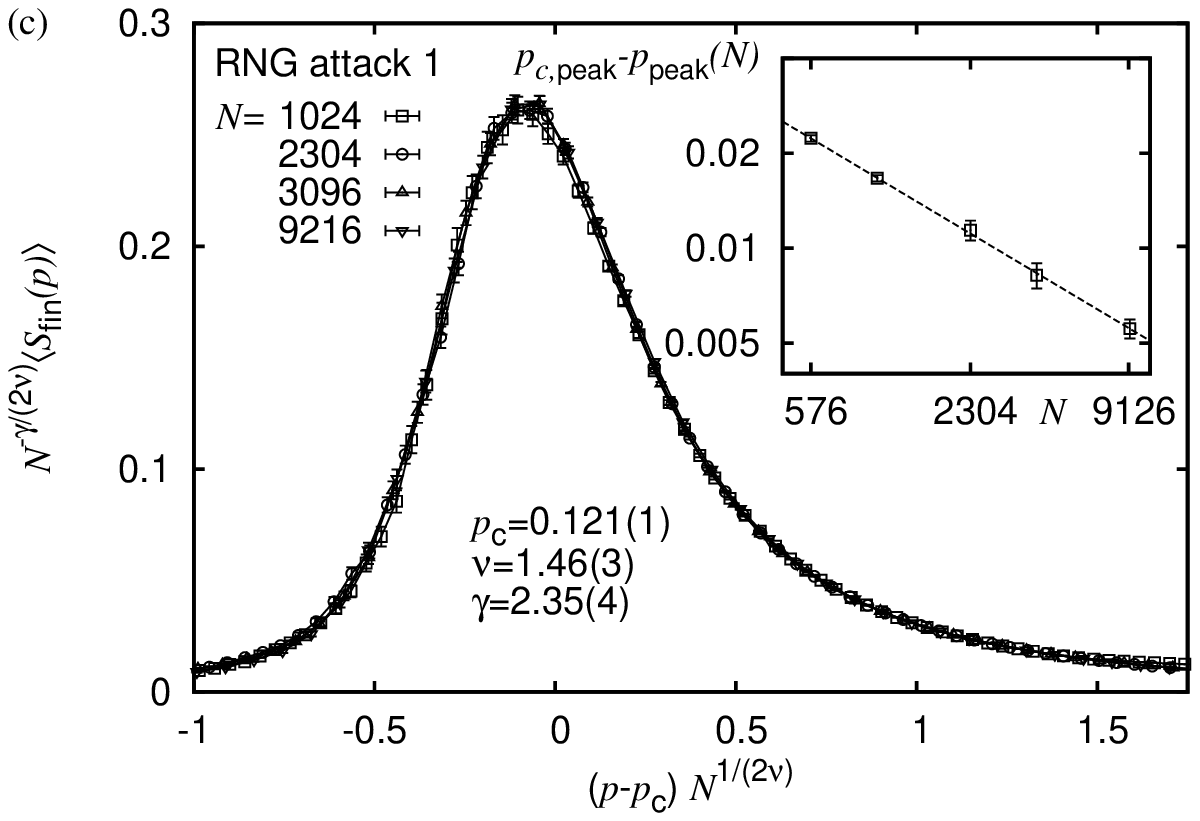}
\end{center}
\caption{Finite-size scaling analysis for the RNG proximity graphs.
(a) order parameter
 for RNGs subject to the three node removal strategies
discussed in Sec.\ \ref{sect:strat}.
(b) the main plot shows the best data collapse of the order parameter 
obtained for a 
degree-based node removal strategy, the inset illustrates the scaling of 
the area under curve for a centrality-based node removal strategy.
The fact that the area seems to converge to zero
is compatible with a critical point $p_c=0$.
(c) the main plot shows the best data collapse for the average size of the 
finite clusters considering a degree-based node removal strategy, the inset
shows the scaling of the peak position. 
\label{fig:analysisRNG}} 
\end{figure}  

\paragraph{Order parameter:}
As first observable we consider $s_{\rm max}$, i.e.\ the relative size 
of the largest cluster of connected nodes. Averaged over different instances 
of, say, size $N$, at a given value of $p$ this yields the \emph{order parameter} 
\begin{eqnarray}
\langle P(p) \rangle=\langle s_{\rm max}(p) \rangle. \label{eq:orderPar}
\end{eqnarray}
This observable scales according to Eq.\ (\ref{eq:scalingAssumption}), where 
$b=\beta/\nu$ and $\beta$ is the order-parameter exponent.
 The data curves for the RNG proximity graphs for all three types
of node removal strategies are shown in Fig.\ \ref{fig:analysisRNG}(a).

For the RNG and GG, the \emph{random failure} node removal strategy simply
corresponds to ordinary random percolation. An extended study of site and bond
percolation for the RNG type proximity graphs can be found in Ref.\
\cite{melchert2013} and in Ref.\  \cite{norrenbrock2014} 
for the GG type proximity graphs, respectively. 
However, note that in these articles $p$ signifies the
fraction of occupied bonds/nodes as opposed to the fraction of deleted nodes.
The respective values of $p_c$ are listed in Tab.\ \ref{tab:critpoints}.
It is apparent, that in the order RNG, GG, DT and MR, the
graphs become less and less susceptible to 
fragment under random node removal. This
correlates well to the average degree $d_{\rm RNG} < d_{\rm
GG} < d_{\rm DT} < d_{\rm eff,MR}(N)$.

Regarding the \emph{degree-based attack strategy} for the RNGs we found that
the best data collapse (obtained for the three system sizes $N=2304, 4096,
9126$ in the range $\epsilon \in [-1,1]$) yields $p_c=0.120(1)$, $\nu=1.33(2)$,
and $\beta=0.148(8)$ with a quality $S=3.63$ (see Refs.\
\cite{houdayer2004,autoScale2009}), see Fig.\ \ref{fig:analysisRNG}(b).  Note
that the numerical values of the critical exponents match the expected values
for 2D percolation, i.e.\ $\nu=4/3\approx 1.333$ and $\beta=5/36\approx0.139$,
quite well. 
Restricting the data analysis to the slightly smaller interval $\epsilon\in
[-0.65,0.65]$, enclosing the critical point on the rescaled $p$-axis, the
optimal scaling parameters are found to be $p_c=0.119(1)$, $\nu=1.41(5)$, and
$\beta=0.14(1)$ with a quality $S=0.98$. Further, fixing $\nu$ and $\beta$ to
their exact values, thus leaving only one parameter to adjust, yields
$p_c=0.119(1)$ with a data-collapse quality $S=3.16$. Hence, for RNGs subject
to a degree-based attack strategy, a fraction of $p_c=0.119(1)$ seems to
suffice in order to decompose the graph instance into small clusters.  Note
that this is already significantly smaller than the above value found for the
case of random node failures. 

The analysis for the proximity graph types GG and DT for the above two
node-removal strategies (i.e.\ random failure and degree-based attack) were
carried out in similar fashion. For the DT ensemble, considering the
degree-based node-removal strategy, the scaling parameters obtained by the FSS
analysis read $p_c=0.377(1)$, $\nu=1.31(7)$, and $0.14(2)$ with a data-collapse
quality $S=0.89$ (obtained for the three system sizes $N=2304, 4096, 9126$ in
the range $\epsilon \in [-0.5,0.75]$).  For comparison: the critical point for
the random node removal strategy is known to be $p_c=0.5$; from our simulated
data we find $p_c=0.500(2)$, $\nu=1.35(13)$, and $\beta=0.13(2)$ with a quality
$S=0.94$ (similar system sizes as above, only in the range $\epsilon \in
[-0.25,0.25]$).

For the case of the GG graphs, we found $p_c=0.263(1)$, $\nu=1.33(4)$ and
$\beta=1.4(2)$ in respect to the degree-based node-removal strategy (obtained
for the system sizes $N=2304, 4096, 9126$ in the range $\epsilon\in [-0.2,0.7]$
with quality $S=1.08$).

However, note that for the geometric MR graphs, an analysis of the 
order parameter following a scaling assumption of the form of 
Eq.\ \ref{eq:scalingAssumption} did not lead to any conclusive results. 
I.e.\ the data curves did not give a satisfactory data collapse. 
Nevertheless, based on the analysis of the fluctuations of the order
parameter, we were able to obtain estimates for the critical point, see
below. In summary, 
as obvious from Tab.\ \ref{tab:critpoints}, degree-based attacks are
more severe than random removals. Again, the resilience
against attacks correlates well with the average degree.

Considering the \emph{centrality-based attack strategy} for the RNGs, we start
out with a more simplified initial analysis. As evident from Fig.\
\ref{fig:analysisRNG}(a), the data curves that describe the scaling of the
order parameter for this setup drop to zero at rather small values of $p$.
Thus, a FSS analysis (as carried out above) comes along with several
difficulties (related to the accessibility of data points in the critical
scaling window). Hence, we first determine the area $A(N)$ under the
order-parameter curves and assess its scaling behavior with increasing system
size $N$ to see whether it converges to a finite value at all. From a fit to
the function $A(N)=a(N+\Delta N)^{-b}$ we find $a=0.20(1)$, $\Delta N=30(4)$,
and $b=0.315(3)$ (reduced chi square $\chi_{\rm red}^2=0.34$; see inset of Fig.\
\ref{fig:analysisRNG}(b)), indicating that indeed $A(N\to\infty)\to0$. If we
neglect the smallest system, we find that a pure power law $A(N)=0.270(3)
N^{-0.302(2)}$ fits the data well (reduced chi-square $\chi_{\rm red}^2=0.65$).
From this we conclude that for RNGs, subject to a centrality-based attack
strategy one has $p_c=0$.  Following this procedure, we also found that under
this attack strategy $p_c=0$ holds true for GGs, DTs, and MRs. Thus,
due to its propensity to fragment graphs at negligible values of
$p$, this strategy is much more efficient than the degree-based strategy,
independent of the type of graph.

\paragraph{Average size of the finite clusters}
As second observable we consider the average size $\langle S_{\rm fin}
(p)\rangle$ of all finite clusters for a particular graph instance, averaged
over different graph instances.  The definition of this observable reads
\cite{Stauffer1992}
\begin{eqnarray}
S_{\rm fin}(p) = \frac{\sum_{s}^\prime s^2\, n_s(p)}{\sum_{s}^\prime s\, n_s(p)}, \label{eq:finClusters}
\end{eqnarray}
where $n_s(p)$ signifies the probability mass function of cluster sizes for a
single graph instance at a given value of $p$. The prime indicates that the
sums run over all clusters excluding the largest cluster for each
graph instance.  The average size of all finite clusters is expected to scale
according to Eq.\ \ref{eq:scalingAssumption}, where $b=-\gamma/\nu$. Therein,
for 2D percolation, the critical exponent $\gamma$ assumes a value of
$\gamma=43/18\approx 2.389$.

\begin{figure}[t!]
\begin{center}
\includegraphics[width=1.0\linewidth]{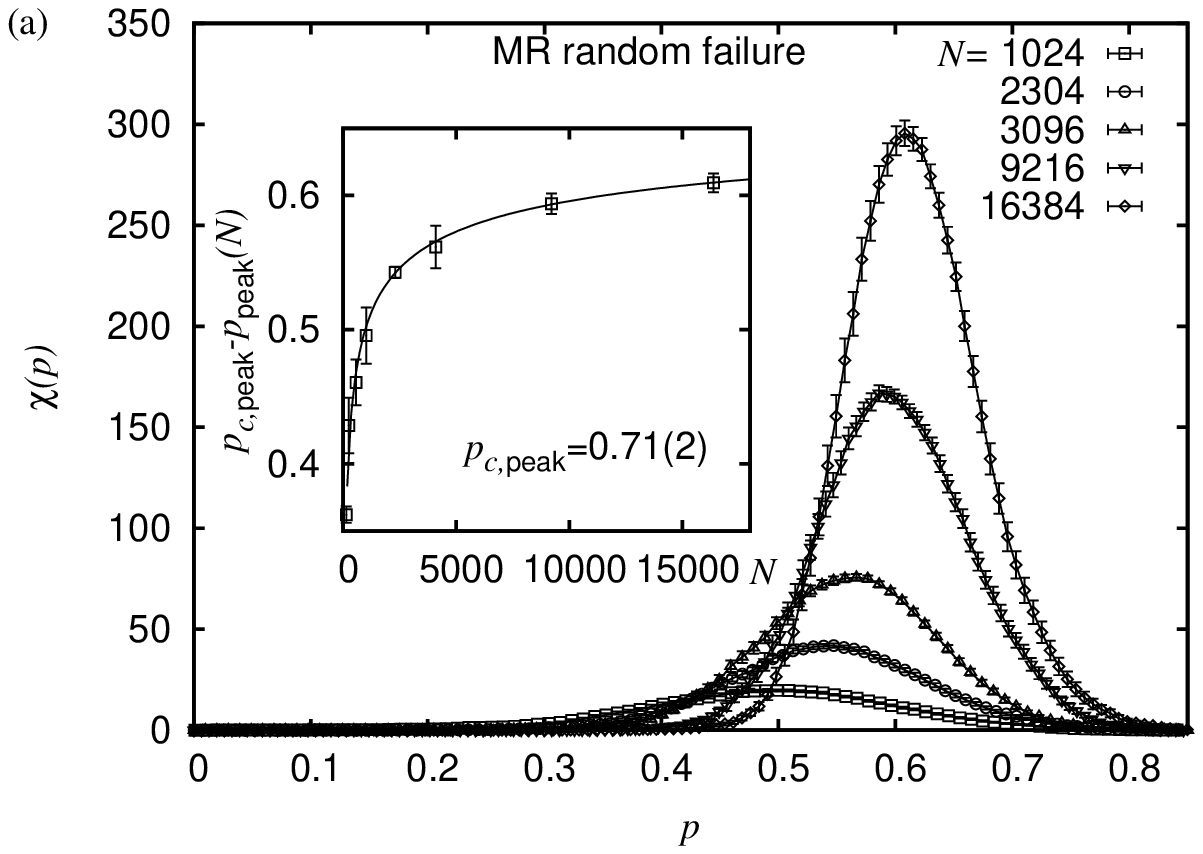}
\includegraphics[width=1.0\linewidth]{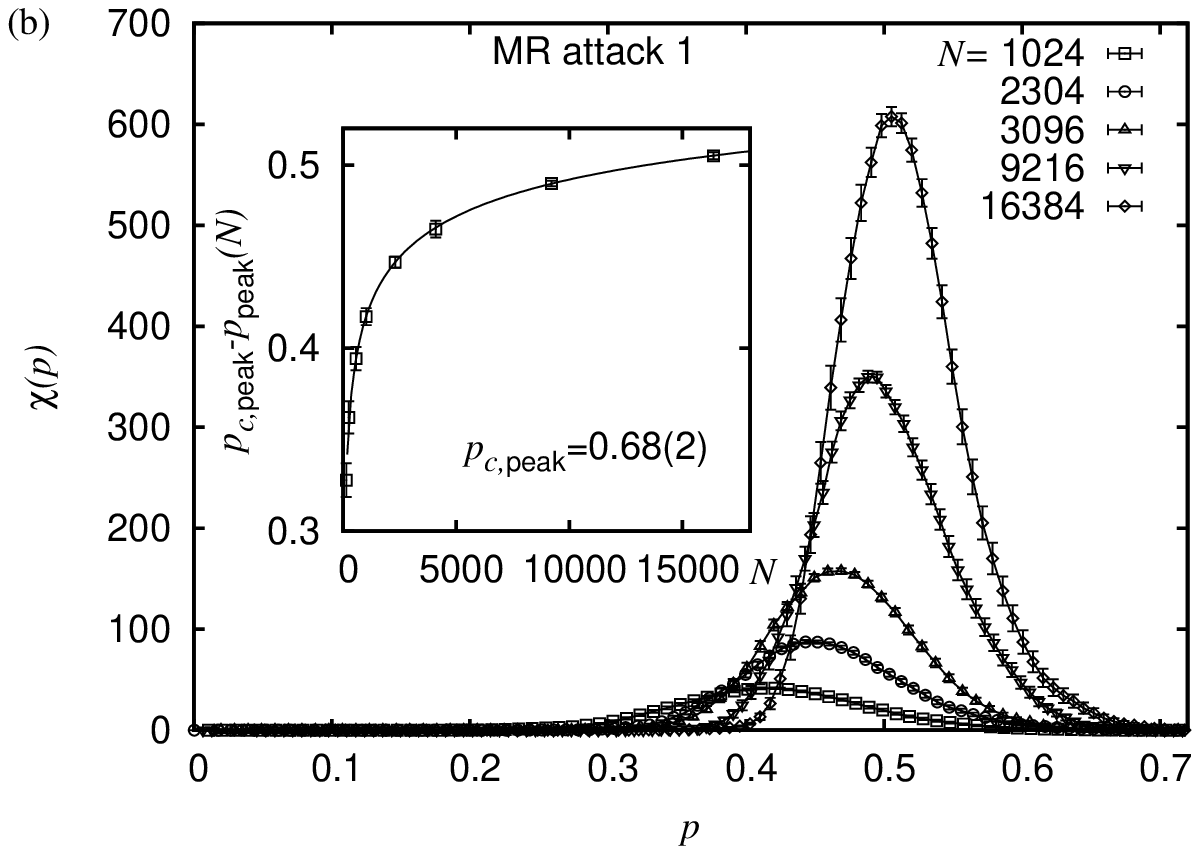}
\end{center}
\caption{Finite-size scaling analysis for the peak position of the finite-size susceptibility curves for the 
MR geometric graphs.
(a) the main plot shows the data curves for the random node removal strategy and the inset illustrates the finite size scaling of the respective 
peak positions (see text for details).
(b) the main plot shows the data curves for the degree-based node removal strategy and the inset illustrates the finite size scaling of the respective 
peak positions (see text for details).
\label{fig:MRsuscept}} 
\end{figure}  

Again, a detailed analysis of this observable for random percolation,
which is equivalent to the random failure node-removal strategy, can be found in
Ref.\ \cite{melchert2013} regarding RNGs, and in Ref.\ \cite{norrenbrock2014}
 with respect to GGs. 

Regarding the \emph{degree-based} attack strategy for RNGs, considering systems
of size $N=1024,2304,4096,9216$ and restricting the data analysis to the interval $\epsilon\in
[-0.5,0.5]$ on the rescaled $p$-axis, the optimal scaling parameters are found
to be $p_c=0.120(1)$, $\nu=1.46(3)$, and $\gamma=2.35(4)$ with a collapse
quality $S=0.91$, see Fig.\ \ref{fig:analysisRNG}(c). Note that here the
estimated value of $\nu$ appears to overestimate the expected value somewhat. 
Apart from that, the numerical
values of the extracted exponents are in reasonable agreement with their
expected values and the estimate of the critical threshold $p_c$ is consistent
with the numerical value found from an analysis of the order parameter.

In addition to the full FSS analysis, we also performed a scaling analysis for
the effective critical points $p_{\rm peak}(N)$ at which the curves of $S_{\rm
fin}$ assume their maximum. Therefore, polynomials of $5$th order were fitted
to the data curves at different system sizes $N$ in order to obtain an estimate
$p_{{\rm peak},i}(N)$ of the peak position. Thereby, the index $i$ labels
independent estimates of the peak position as obtained by bootstrap
resampling. For the analysis, we considered $20$ bootstrap data sets, e.g.\
resulting in the estimate $p_{\rm peak}(N=9216)=0.1057(4)$ for the RNG
regarding the degree-based attack strategy. Considering systems of size $N>500$ and
assuming the scaling form 
\begin{eqnarray}
p_{\rm peak}(N) = p_{c,{\rm peak}} - a N ^{-b}, \label{eq:avSizeFin_fitPeak}
\end{eqnarray}
we yield the fit parameters $p_{c,{\rm peak}}=0.111(1)$, $b=0.50(3)$ and $a=O(1)$ for a
reduced chi-square $\chi^2_{\rm red}=0.08$, see inset of Fig.\
\ref{fig:analysisRNG}(c). 
This result indicates that the peak seems to be positioned off criticality
at a value slightly below $p_c$, cf.\ Fig.\ \ref{fig:analysisRNG}(c).
However, including also very small systems we yield $p_{c,{\rm peak}}=0.119(5)$, $b=0.29(6)$ and $a=O(1)$ for a
reduced chi-square $\chi^2_{\rm red}=3.89$, in good agreement with the value of $p_c$ obtained 
from an analysis of the order parameter.
 Following this procedure by considering RNGs subject
to a random node failure we yield $p_c=0.196(8)$, which compares well to estimate
obtained from an analysis of the order parameter (see Tab.\
\ref{tab:critpoints}).
An analysis of the peak positions for all other types of proximity graphs led to qualitatively similar results.
Hence, we do not elaborate on them here.

Whenever we analyzed the order parameter, we also analyzed the respective
fluctuations, giving rise to the finite-size susceptibility $\chi(p)$
\begin{eqnarray}
\chi(p)= N [ \langle s_{\rm max}^2(p) \rangle - \langle s_{\rm max}(p) \rangle^2  ]. \label{eq:suscept}
\end{eqnarray}
These curves also feature a pronounced peak and exhibit the same scaling
behavior as the average size of the finite clusters discussed above. Here, we
also performed a scaling analysis of the peak positions of the $\chi(p)$
curves, similar to that performed for the peaks of the previous observable.
Albeit this did not lead to new insight for the various types of proximity
graphs, it was a valuable method to estimate critical points for the MR
geometric graphs. In this regard, for MRs subject to a random node removal we
find $p_c=0.71(2)$, see Fig.\ \ref{fig:MRsuscept}(a). Further, for MRs subject to the degree-based node removal
strategy we obtain $p_c=0.68(2)$, see Fig.\ \ref{fig:MRsuscept}(b). Hence, for the MRs we cannot rule out that
the estimates for both critical points actually agree within error-bars. This
might be attributed to the rather high degree of the individual nodes, and, from
a statistical point of view, the extensive overlap of the individual
node-neighborhoods within the range of the underlying ``connectivity radius''.
Hence, due to the high number of redundant node-to-node paths which easily
allow to compensate for deleted nodes, the effect caused by the removal of a
randomly chosen node does not differ much from the effect caused by the removal
of a node with a particularly large degree.

\subsubsection{Analysis of the hop diameter:}\label{subsect:hopDiam}

\begin{figure}[t!]
\begin{center}
\includegraphics[width=1.0\linewidth]{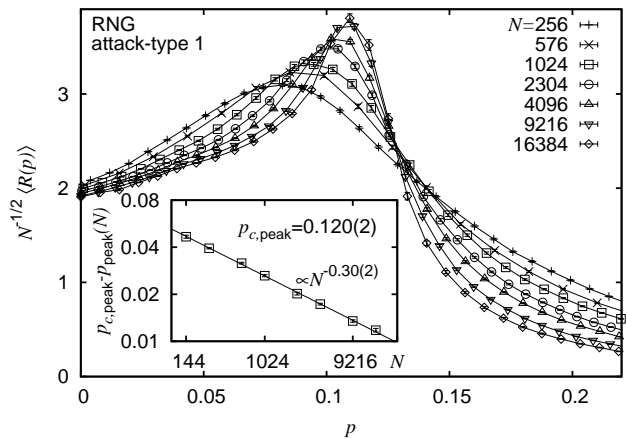}
\end{center}
\caption{Finite-size scaling analysis for the peak position of the diameter curves for the 
RNG proximity graphs subject to a degree-based attack strategy.
The main plot shows the diameter of the graphs as function of the removed fraction
of nodes following the attack strategy ``attack 1'' (discussed in Sec.\ \ref{sect:strat}).
The inset illustrates the scaling behavior of the effective peak position $p_{{\rm peak}}(N)$
as function of the system size $N$.
\label{fig:RNGmaxLen}} 
\end{figure}  

The last observable, studied in the context of the fragmentation analysis is
related to the diameter $R(p)$ of the graphs as function of the fraction $p$ of
removed nodes.  Here, the diameter of a graph indicates the longest among all
finite shortest paths.  In Fig.\ \ref{fig:RNGmaxLen}, the data curves of the
diameter for the particular choice of RNG proximity graphs subject to a
degree-based node removal strategy are shown.  For the particular case of
non-fragmented RNGs, i.e.\ at $p=0$, the diameter (averaged over different
realizations of point-sets) was previously found to scale as $\langle R\rangle\propto
N^{1/2}$ \cite{melchert2013}. In view of these prior results, the data 
curves in Fig.\ \ref{fig:RNGmaxLen} are scaled so as to assume a fixed value
at $p=0$.
As evident from the figure the data curves assume a peak value when a certain fraction of nodes is removed. 
This appears to be quite intuitive: if nodes are removed from one of the graph instances introduced in
Sec.\ \ref{sect:proxigraph}, redundant edges will disappear (on average) resulting
in an increasing node-to-node distance. As soon as the value of $p$ exceeds the
percolation threshold of the respective setup (i.e.\ graph type and node removal strategy), 
the graph instance decomposes into several ``small'' clusters accompanied by a decreasing
node-to-node distance.  With increasing system size $N$, the position $p_{\rm
peak}(N)$ of the peak shifts towards larger values of $p$. For RNGs subject to
a degree-based node removal, a fit-function of the form similar to Eq.\ \ref{eq:avSizeFin_fitPeak} 
yields the fit parameters $p_c=0.120(2)$,
$a=O(10^{-1})$ and $b=0.30(2)$ ($\chi^2_{\rm red}=0.90$). Similarly, for DTs we
find $p_c=0.378(4)$, $a=O(10^{-2})$ and $b=0.3(2)$ ($\chi^2_{\rm red}=0.14$).
The resulting asymptotic peak positions are in good agreement with the value
obtained from a FSS analysis of the order parameter, cf.\ Tab.\ \ref{tab:critpoints}.
For the case of a random node failure, the results obtained from the scaling of
the peak position fits the results from the order parameter analysis similarly
well. E.g., for the case of RNGs we find $p_c=0.199(2)$, cf.\ Tab.\
\ref{tab:critpoints}.  Albeit we performed a similar analyses for GGs and DTs,
resulting in qualitatively similar results, we do not elaborate on them here.
No analyses were performed for the centrality-based node removal strategy.

\subsection{$N-1$ resilience}\label{subsect:stability} 

The actual most important application example regarding proximity graphs are
wireless ad hoc networks.  Nevertheless, there might be some other fields of
application for them.  Proximity graphs ensure connectivity and the total
length of all involved edges is small in comparison to many other networks that
feature this quality.  For applications where edges are expensive and
connectivity is crucial, the topology of proximity graphs might be a good
candidate to install.  Up to here, we have assumed that the capacities of the
edges and nodes are infinitely large, so the network components do not
overload, regardless how intensive they get strained. 
In real scenarios, if some nodes or edges malfunction, network
components which are hardly used under normal circumstances might become
essential at once.  In consequence, since the hardly used components are not
designed to handle such a burden, this might trigger a cascading breakdown of
the whole network \cite{Bao2009,Crucitti2004-2,Motter2004,Motter2002}.
Therefore, it is reasonable to equip all network components with sufficient
capacity.

To ensure that the network operates orderly under all circumstances when one
component drops out, referred to as ``$N-1$ resilience'' (or $N-1$ stability,
or $N-1$ criterion),
it is necessary to know the most adverse scenario that can happen.
When having a transport model in mind, where some quantities have to be
transported between all pairs of nodes,
\cite{Motter2002,Motter2004,Bao2009}
 the betweenness centrality \cite{newman2001shortest-path} is a measure
of the capacity each node or edge has to provide in a well functioning
situation. When one node or edge fails, given that the network is still
connected, the loads have to be redistributed, visible from a recalculation of
the betweenness centrality. 
In some nodes or edges the centrality will
increase \cite{ouyang2014,Hartmann2014}, 
corresponding to a higher capacity these nodes or edges have to
provide a priori.  The value of the highest increment, which is called backup
capacity \cite{Hartmann2014} $\Delta b_{\rm node}$ or $\Delta b_{\rm edge}$,
provides an estimate for the additional costs for each node or edge
that must be invested to protect the network against cascading failures upon
such an incident.

\begin{figure}[bt]
\centerline{
\includegraphics[width=0.9\linewidth]{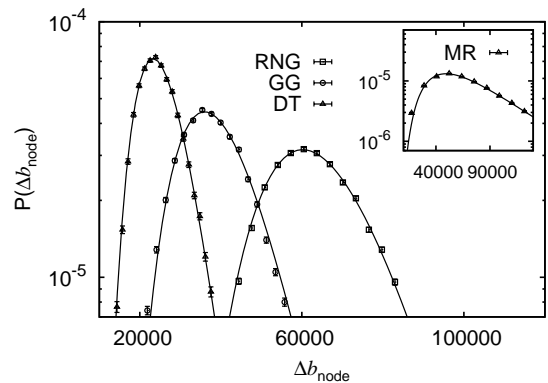}}
\caption{
Probability mass function of the backup capacity, which is needed
to ensure $N-1$ stability (see text) 
$\Delta b_{\rm node}$ for different network topologies 
(initial system size: $N\!=\!1024$).
The probability mass function concerning this measure has been made 
by analyzing $40000$ realizations of the disorder.
The data has been fitted by a log-normal distribution with moderate quality 
(reduced $\chi_{\rm red}^2$ between $0.65$ and $4.46$).
\label{fig:dij_node}}
\end{figure}

\begin{figure}[bt]
\centerline{
\includegraphics[width=0.9\linewidth]{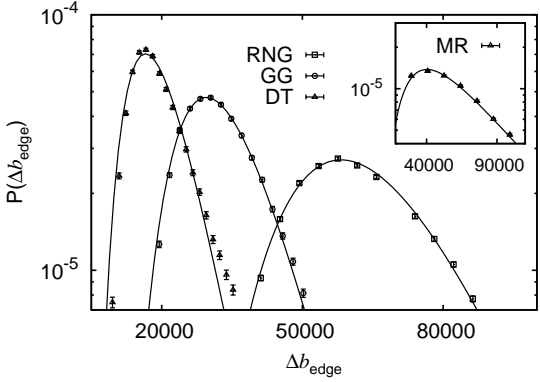}}
\caption{
Probability mass function of $\Delta b_{\rm edge}$ for different network
topologies (initial system size: $N\!=\!1024$).  Removing the edge featuring
the largest betweenness centrality value, the largest increment of the
betweenness centrality of the other edges $\Delta b_{\rm edge}$ has been
monitored.  The probability mass function concerning this measure has been made
by analyzing $40000$ realizations of the disorder.  The data has been fitted by
a log-normal distribution with good quality for GG (reduced $\chi_{\rm
red}^2\!=\!0.26$) and RNG (reduced $\chi_{\rm red}^2\!=\!2.71$), but poor
quality for MR (reduced $\chi_{\rm red}^2\!=\!5.2$) and DT (reduced $\chi_{\rm
red}^2\!=\!12.2$).
\label{fig:dij_edge}}
\end{figure}

\begin{figure}[bt]
\centerline{
\includegraphics[width=1\linewidth]{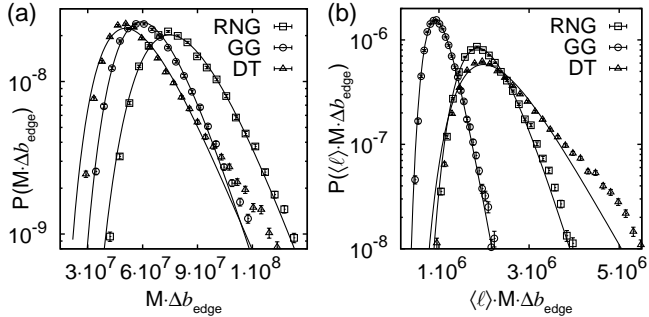}}
\caption{
Probability mass functions of (a) $M\cdot\Delta b_{\rm edge}$ and (b) $\langle\ell\rangle\cdot M\cdot\Delta b_{\rm edge}$ for different network
topologies at $N\!=\!1024$. $M$ describes the number of edges in the respective graph and $\langle \ell\rangle$ denotes their respective edge-length mean value.
The data has been fitted by a log-normal distribution with predominant low quality (regarding (a): reduced $\chi_{\rm
red}^2\!=\!6.5$ for GG data, reduced $\chi_{\rm red}^2\!=\!4.99$ for RNG data and reduced $\chi_{\rm red}^2\!=\!12.75$ for DT, regarding (b): reduced $\chi_{\rm
red}^2\!=\!2.9$ for GG data, reduced $\chi_{\rm red}^2\!=\!5.34$ for RNG data and reduced $\chi_{\rm red}^2\!=\!20.09$ for DT)
\label{fig:weightedDist}}
\end{figure}
The betweenness centrality has been calculated based on Dijkstra's algorithm
\cite{Cormen2001}, i.e.\ the edge lengths have been taken into account for
calculating the shortest path.  The resulting probability mass functions of
$\Delta b_{\rm node}$ and $\Delta b_{\rm edge}$ 
are illustrated in Fig.\ \ref{fig:dij_node} and \ref{fig:dij_edge}, 
respectively, for different network ensembles.

For the case of $\Delta b_{\rm edge}$ however,
 using a breadth-first search, i.e., without taking the actual edge
lengths into account, the probability mass functions look qualitatively almost
identical to the case of $\Delta b_{\rm node}$
without notable distinctions. From the statistics mediated by
the figures it is evident that the \emph{typical} 
backup capacity of DT networks is smallest while of RNG networks
it is the highest.
This means that the structure of the network of RNG networks
is more vulnerable such that one
has to invest more into the capacity of the edges in order to ensure
$N-1$ stability. This is not surprising, since the RNG is a sub-graph of the
GG and DT and includes less edges.  
On the other hand, due to the lack of the additional 
edges of the RNG in comparison to the others,
the investment to provide the backup capacity 
must be applied for less edges. Thus,
it makes sense to ask for the total backup capacity, either per edge,
if investments cost are dominated by the number of connections,
or per unit length, if investments are dominated by the length
of the edges.
It becomes evident from Fig.\ \ref{fig:weightedDist}(a) 
that for the former case,
the \emph{typical} total investment ($M\cdot\Delta b$) is relatively
speaking still the same for all three ensembles. For the second case, i.e.,
taking also edge lengths into account 
(Fig.\ \ref{fig:weightedDist}(b)), it turns out that the investment of the 
DT is about the same level as the RNG. The GG appears to be 
the most cost efficient graph if this scenario is at hand.

\subsection{Networks with same total length}\label{subsect:costs} 

\begin{figure}[bt]
\centerline{
\includegraphics[width=0.9\linewidth]{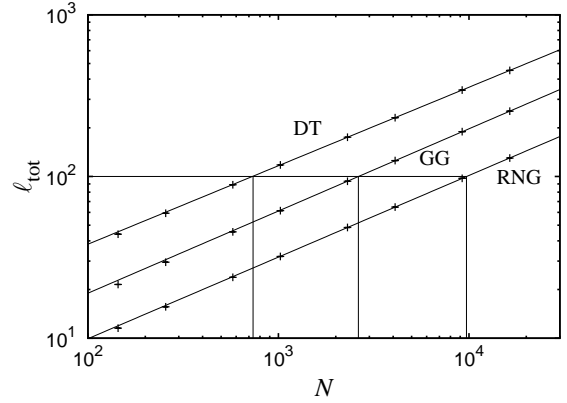}}
\caption{
The scaling behavior of the total edge length $\ell_{\rm tot}$ of the RNG, GG
and DT.  Each data point has been created by averaging over $2000$ instances.
In each case the total edge length seems to scale according to $\ell_{\rm
tot}\sim N^{0.5}$ for large systems.  We used the fit function $\ell_{\rm tot}
= a(N+b)^c$. By taking the system sizes $N=2304,4096,9216,16384$ into account,
we found $a=4.02(2)$, $b=19(5)$, $c=0.4868(4)$ for the DT ($\chi_{\rm red}^2$ =
2.44), $a=1.91(1)$, $b=-47(5)$, $c=0.5039(5)$ for the GG ($\chi_{\rm red}^2 =
1.31$) and $a=0.99(1)$, $b=-20(3)$, $c=0.5024(3)$ for RNG ($\chi_{\rm red}^2 =
0.46$).  \label{fig:sumLen}} \end{figure}

To compensate for the simple resilience effect created by simply
exhibiting more edges, we also compared the different topologies of the
proximity graphs featuring the same total edge length $\ell_{\rm tot}$.
Therefore, we measured the scaling behavior of this quantity for the different
proximity graph types (see Fig.\ \ref{fig:sumLen}).  The figure provides the
number of nodes which have to be added to the RNG and GG in order to get same
total edge length as the respective DT.  E.g., it is evident from the figure
that a DT with $N=718$ nodes has the same total edge length $\ell_{\rm tot} =
100$ on average as a GG with $N=2625$ and a RNG with $N=9783$ nodes.  Since the
fragmentation thresholds for the different node-removal strategies are known
(Tab.\ \ref{tab:critpoints}), it can be calculated easily ($N\cdot p_c$) for
each topology how many nodes must be removed until the respective network
decomposes into small clusters.  E.g., if $\ell_{\rm tot} = 100$, the RNG will tolerate $2005$
randomly removed nodes. In contrast, the GG tolerates $958$ and the DT
tolerates merely $359$ nodes that fail randomly.
 As a consequence, implementing the
topology of the RNG will be the most reasonable, if installing edges is much
more expensive than adding further nodes.  Certainly, the additional edges of
the DT and GG increases the stability, but the benefit of those is small in
comparison to the edges which are contained in the RNG anyway.

\section{Conclusion\label{sect:conclusion}}
In the presented article, the robustness of three types of proximity graphs and
a particular geometric random graph (see Sec.\ \ref{sect:proxigraph}), i.e.\
their ability to function well even if they are subject to random failures and
targeted attacks, was put under scrutiny.  For this purpose we generated
instances of the considered graph types and successively removed nodes
according to three different node removal strategies (see Sec.\
\ref{sect:strat}). Once the fraction of removed nodes exceeds a certain
threshold (characteristic for the particular graph type and node deletion
scheme), the underlying graph instance decomposes into many small clusters.
Using standard observables from percolation theory (see Sec.\
\ref{subsect:fragmentationProcedure}), the critical node removal thresholds
were determined for the different graph types and deletion strategies, see
Tab.\ \ref{tab:critpoints}. Therein, so as to yield maximally justifiable results through numerical redundancy, 
we considered various observables to estimate the critical points and exponents.
In order of increasing severity, these strategies have an intuitive order: a
random node removal mechanism, equivalent to ordinary random percolation, is
less severe than a degree-based node removal strategy which takes into account
particular node-related local details (i.e.\ the node-degrees) to optimize the
order of node removals during the fragmentation procedure. As evident from
Tab.\ \ref{tab:critpoints}, both removal schemes result in finite critical
points.  The latter strategy is again less severe than the centrality-based
node removal mechanism, which takes into account global information (i.e.\ the
set of shortest paths that connect all pairs of nodes) which is used to impose
a maximally efficient structural damage by preferentially removing nodes with
maximal betweenness centrality (i.e.\ the most relevant nodes).  As evident
from Tab.\ \ref{tab:critpoints} and the discussion in Sec.\
\ref{subsect:fragmentationProcedure}, the latter node removal scheme requires
to delete only 
a negligible amount of nodes until the graph decomposes into small
clusters.
A peculiar result are the fragmentation thresholds related to the random failure
and degree-based node removal for the MR geometric graph. As discussed in 
Sec.\ \ref{subsect:fragmentationProcedure} we cannot rule out that
the estimates for both critical points agree within error-bars. This
might be attributed to the extensive overlap of the individual
node-neighborhoods within the range of the underlying ``connectivity radius''.
Hence, due to the high number of redundant node-to-node paths which easily
allow to compensate for deleted nodes, the effect caused by the removal of a
randomly chosen node does not differ much from the effect caused by the removal
of a node with a particularly large degree. 

For a given node removal strategy, the sequence of critical points for the
sub-graph hierarchy ${\rm RNG} \subset {\rm GG} \subset {\rm DT}$ follow the
commonly accepted belief that the percolation threshold (or here: the
fragmentation threshold) is a non decreasing function of the average degree.
This is in full accord with the containment principle due to Fisher
\cite{Fisher1961}, stating that if $G^\prime$ is a sub-graph of $G$, then it
holds that $p_c^{G^\prime} \leq p_c^G$ for both, bond and site percolation.

Finally, we considered the backup capacity, which is the largest
betweenness-centrality increment of the nodes (or edges) after removing the
most important node (or edge) beforehand, for the different graph types.
Thus, via sufficient backup a graph is made $N-1$ resilient.
Regarding the three studied proximity graph ensembles, it turned out that the 
DT is the
most cost efficient one assuming that the backup investments
are dominated by improving
the nodes. On the other hand, if one has to backup the edges,
the more cost
efficient one will be either DT or GG, depending on whether
the investment depends mainly on the number or on the length of
the edges.

For further studies, it would be very interesting to
evaluate these simple spatial planar ensembles in the context of
more complex transportation networks, like for
steady state power grids in the power-flow approximation, \cite{kundur1994}
or for networks of truly dynamically coupled oscillators
as for Kurmatoto-like models\cite{filatrella2008,rohden2012,witthaut2012}.

\begin{acknowledgments}
CN gratefully acknowledges financial support from the Lower Saxony 
research network ``Smart Nord'' which acknowledges the support of 
the Lower Saxony Ministry of Science and Culture through the 
``Nieders\"achsisches Vorab'' grant program (grant ZN 2764 / ZN 2896). 
OM gratefully acknowledges financial support from the DFG 
(\emph{Deutsche Forschungsgemeinschaft}) under grant HA3169/3-1.
The simulations were performed at the HPC Cluster HERO, located at 
the University of Oldenburg (Germany) and funded by the DFG through
its Major Instrumentation Programme (INST 184/108-1 FUGG) and the
Ministry of Science and Culture (MWK) of the Lower Saxony State.
\end{acknowledgments}

\bibliography{research.bib}

\begin{thebibliography}{57}
\expandafter\ifx\csname natexlab\endcsname\relax\def\natexlab#1{#1}\fi
\expandafter\ifx\csname bibnamefont\endcsname\relax
  \def\bibnamefont#1{#1}\fi
\expandafter\ifx\csname bibfnamefont\endcsname\relax
  \def\bibfnamefont#1{#1}\fi
\expandafter\ifx\csname citenamefont\endcsname\relax
  \def\citenamefont#1{#1}\fi
\expandafter\ifx\csname url\endcsname\relax
  \def\url#1{\texttt{#1}}\fi
\expandafter\ifx\csname urlprefix\endcsname\relax\def\urlprefix{URL }\fi
\providecommand{\bibinfo}[2]{#2}
\providecommand{\eprint}[2][]{\url{#2}}

\bibitem[{\citenamefont{Stauffer}(1979)}]{stauffer1979}
\bibinfo{author}{\bibfnamefont{D.}~\bibnamefont{Stauffer}},
  \bibinfo{journal}{Phys. Rep.} \textbf{\bibinfo{volume}{54}},
  \bibinfo{pages}{1} (\bibinfo{year}{1979}).

\bibitem[{\citenamefont{Stauffer and Aharony}(1992)}]{Stauffer1992}
\bibinfo{author}{\bibfnamefont{D.}~\bibnamefont{Stauffer}} \bibnamefont{and}
  \bibinfo{author}{\bibfnamefont{A.}~\bibnamefont{Aharony}},
  \emph{\bibinfo{title}{Introduction to Percolation Theory}}
  (\bibinfo{publisher}{Taylor \& Francis}, \bibinfo{year}{1992}).

\bibitem[{\citenamefont{Albert et~al.}(2004)\citenamefont{Albert, Albert, and
  Nakarado}}]{Albert2004}
\bibinfo{author}{\bibfnamefont{R.}~\bibnamefont{Albert}},
  \bibinfo{author}{\bibfnamefont{I.}~\bibnamefont{Albert}}, \bibnamefont{and}
  \bibinfo{author}{\bibfnamefont{G.~L.} \bibnamefont{Nakarado}},
  \bibinfo{journal}{Phys. Rev. E} \textbf{\bibinfo{volume}{69}},
  \bibinfo{pages}{025103} (\bibinfo{year}{2004}).

\bibitem[{\citenamefont{Jiang and Claramunt}(2004)}]{Jiang2004}
\bibinfo{author}{\bibfnamefont{B.}~\bibnamefont{Jiang}} \bibnamefont{and}
  \bibinfo{author}{\bibfnamefont{C.}~\bibnamefont{Claramunt}},
  \bibinfo{journal}{GeoInformatica} \textbf{\bibinfo{volume}{8}},
  \bibinfo{pages}{157} (\bibinfo{year}{2004}).

\bibitem[{\citenamefont{Choi et~al.}(2006)\citenamefont{Choi, Barnett, and
  Chon}}]{Choi2006}
\bibinfo{author}{\bibfnamefont{J.~H.} \bibnamefont{Choi}},
  \bibinfo{author}{\bibfnamefont{G.~A.} \bibnamefont{Barnett}},
  \bibnamefont{and} \bibinfo{author}{\bibfnamefont{B.-S.} \bibnamefont{Chon}},
  \bibinfo{journal}{Global Networks} \textbf{\bibinfo{volume}{6}},
  \bibinfo{pages}{81} (\bibinfo{year}{2006}).

\bibitem[{\citenamefont{Barab{\'a}si et~al.}(2000)\citenamefont{Barab{\'a}si,
  Albert, and Jeong}}]{Barabasi2000}
\bibinfo{author}{\bibfnamefont{A.-L.} \bibnamefont{Barab{\'a}si}},
  \bibinfo{author}{\bibfnamefont{R.}~\bibnamefont{Albert}}, \bibnamefont{and}
  \bibinfo{author}{\bibfnamefont{H.}~\bibnamefont{Jeong}},
  \bibinfo{journal}{Physica A} \textbf{\bibinfo{volume}{281}},
  \bibinfo{pages}{69} (\bibinfo{year}{2000}).

\bibitem[{\citenamefont{Albert et~al.}(2000)\citenamefont{Albert, Jeong, and
  Barabasi}}]{Albert2000}
\bibinfo{author}{\bibfnamefont{R.}~\bibnamefont{Albert}},
  \bibinfo{author}{\bibfnamefont{H.}~\bibnamefont{Jeong}}, \bibnamefont{and}
  \bibinfo{author}{\bibfnamefont{A.-L.} \bibnamefont{Barabasi}},
  \bibinfo{journal}{Nature} \textbf{\bibinfo{volume}{406}},
  \bibinfo{pages}{378} (\bibinfo{year}{2000}).

\bibitem[{\citenamefont{Callaway et~al.}(2000)\citenamefont{Callaway, Newman,
  Strogatz, and Watts}}]{Callaway2000}
\bibinfo{author}{\bibfnamefont{D.~S.} \bibnamefont{Callaway}},
  \bibinfo{author}{\bibfnamefont{M.~E.~J.} \bibnamefont{Newman}},
  \bibinfo{author}{\bibfnamefont{S.~H.} \bibnamefont{Strogatz}},
  \bibnamefont{and} \bibinfo{author}{\bibfnamefont{D.~J.} \bibnamefont{Watts}},
  \bibinfo{journal}{Phys. Rev. Lett.} \textbf{\bibinfo{volume}{85}},
  \bibinfo{pages}{5468} (\bibinfo{year}{2000}).

\bibitem[{\citenamefont{Crucitti
  et~al.}(2004{\natexlab{a}})\citenamefont{Crucitti, Latora, Marchiori, and
  Rapisarda}}]{Crucitti2004-1}
\bibinfo{author}{\bibfnamefont{P.}~\bibnamefont{Crucitti}},
  \bibinfo{author}{\bibfnamefont{V.}~\bibnamefont{Latora}},
  \bibinfo{author}{\bibfnamefont{M.}~\bibnamefont{Marchiori}},
  \bibnamefont{and}
  \bibinfo{author}{\bibfnamefont{A.}~\bibnamefont{Rapisarda}},
  \bibinfo{journal}{Physica A} \textbf{\bibinfo{volume}{340}},
  \bibinfo{pages}{388} (\bibinfo{year}{2004}{\natexlab{a}}).

\bibitem[{\citenamefont{Gallos et~al.}(2005)\citenamefont{Gallos, Cohen,
  Argyrakis, Bunde, and Havlin}}]{Gallos2005}
\bibinfo{author}{\bibfnamefont{L.~K.} \bibnamefont{Gallos}},
  \bibinfo{author}{\bibfnamefont{R.}~\bibnamefont{Cohen}},
  \bibinfo{author}{\bibfnamefont{P.}~\bibnamefont{Argyrakis}},
  \bibinfo{author}{\bibfnamefont{A.}~\bibnamefont{Bunde}}, \bibnamefont{and}
  \bibinfo{author}{\bibfnamefont{S.}~\bibnamefont{Havlin}},
  \bibinfo{journal}{Phys. Rev. Lett.} \textbf{\bibinfo{volume}{94}},
  \bibinfo{pages}{188701} (\bibinfo{year}{2005}).

\bibitem[{\citenamefont{Holme et~al.}(2002)\citenamefont{Holme, Kim, Yoon, and
  Han}}]{Holme2002}
\bibinfo{author}{\bibfnamefont{P.}~\bibnamefont{Holme}},
  \bibinfo{author}{\bibfnamefont{B.~J.} \bibnamefont{Kim}},
  \bibinfo{author}{\bibfnamefont{C.~N.} \bibnamefont{Yoon}}, \bibnamefont{and}
  \bibinfo{author}{\bibfnamefont{S.~K.} \bibnamefont{Han}},
  \bibinfo{journal}{Phys. Rev. E} \textbf{\bibinfo{volume}{65}},
  \bibinfo{pages}{056109} (\bibinfo{year}{2002}).

\bibitem[{\citenamefont{Cohen et~al.}(2001)\citenamefont{Cohen, Erez, ben
  Avraham, and Havlin}}]{Cohen2001}
\bibinfo{author}{\bibfnamefont{R.}~\bibnamefont{Cohen}},
  \bibinfo{author}{\bibfnamefont{K.}~\bibnamefont{Erez}},
  \bibinfo{author}{\bibfnamefont{D.}~\bibnamefont{ben Avraham}},
  \bibnamefont{and} \bibinfo{author}{\bibfnamefont{S.}~\bibnamefont{Havlin}},
  \bibinfo{journal}{Phys. Rev. Lett.} \textbf{\bibinfo{volume}{86}},
  \bibinfo{pages}{3682} (\bibinfo{year}{2001}).

\bibitem[{\citenamefont{Huang et~al.}(2011)\citenamefont{Huang, Gao, Buldyrev,
  Havlin, and Stanley}}]{Huang2011}
\bibinfo{author}{\bibfnamefont{X.}~\bibnamefont{Huang}},
  \bibinfo{author}{\bibfnamefont{J.}~\bibnamefont{Gao}},
  \bibinfo{author}{\bibfnamefont{S.~V.} \bibnamefont{Buldyrev}},
  \bibinfo{author}{\bibfnamefont{S.}~\bibnamefont{Havlin}}, \bibnamefont{and}
  \bibinfo{author}{\bibfnamefont{H.~E.} \bibnamefont{Stanley}},
  \bibinfo{journal}{Phys. Rev. E} \textbf{\bibinfo{volume}{83}},
  \bibinfo{pages}{065101} (\bibinfo{year}{2011}).

\bibitem[{\citenamefont{Kurant et~al.}(2007)\citenamefont{Kurant, Thiran, and
  Hagmann}}]{Kurant2007}
\bibinfo{author}{\bibfnamefont{M.}~\bibnamefont{Kurant}},
  \bibinfo{author}{\bibfnamefont{P.}~\bibnamefont{Thiran}}, \bibnamefont{and}
  \bibinfo{author}{\bibfnamefont{P.}~\bibnamefont{Hagmann}},
  \bibinfo{journal}{Phys. Rev. E} \textbf{\bibinfo{volume}{76}},
  \bibinfo{pages}{026103} (\bibinfo{year}{2007}).

\bibitem[{\citenamefont{Paul et~al.}(2004)\citenamefont{Paul, Tanizawa, Havlin,
  and Stanley}}]{Paul2004}
\bibinfo{author}{\bibfnamefont{G.}~\bibnamefont{Paul}},
  \bibinfo{author}{\bibfnamefont{T.}~\bibnamefont{Tanizawa}},
  \bibinfo{author}{\bibfnamefont{S.}~\bibnamefont{Havlin}}, \bibnamefont{and}
  \bibinfo{author}{\bibfnamefont{H.~E.} \bibnamefont{Stanley}},
  \bibinfo{journal}{Eur. Phys. J. B} \textbf{\bibinfo{volume}{38}},
  \bibinfo{pages}{187} (\bibinfo{year}{2004}).

\bibitem[{\citenamefont{Shargel et~al.}(2003)\citenamefont{Shargel, Sayama,
  Epstein, and Bar-Yam}}]{Shargel2003}
\bibinfo{author}{\bibfnamefont{B.}~\bibnamefont{Shargel}},
  \bibinfo{author}{\bibfnamefont{H.}~\bibnamefont{Sayama}},
  \bibinfo{author}{\bibfnamefont{I.~R.} \bibnamefont{Epstein}},
  \bibnamefont{and} \bibinfo{author}{\bibfnamefont{Y.}~\bibnamefont{Bar-Yam}},
  \bibinfo{journal}{Phys. Rev. Lett.} \textbf{\bibinfo{volume}{90}},
  \bibinfo{pages}{068701} (\bibinfo{year}{2003}).

\bibitem[{\citenamefont{Tanizawa et~al.}(2012)\citenamefont{Tanizawa, Havlin,
  and Stanley}}]{Tanizawa2012}
\bibinfo{author}{\bibfnamefont{T.}~\bibnamefont{Tanizawa}},
  \bibinfo{author}{\bibfnamefont{S.}~\bibnamefont{Havlin}}, \bibnamefont{and}
  \bibinfo{author}{\bibfnamefont{H.~E.} \bibnamefont{Stanley}},
  \bibinfo{journal}{Phys. Rev. E} \textbf{\bibinfo{volume}{85}},
  \bibinfo{pages}{046109} (\bibinfo{year}{2012}).

\bibitem[{\citenamefont{Brandes}(2008)}]{Brandes2008}
\bibinfo{author}{\bibfnamefont{U.}~\bibnamefont{Brandes}},
  \bibinfo{journal}{Social Networks} \textbf{\bibinfo{volume}{30}},
  \bibinfo{pages}{136} (\bibinfo{year}{2008}).

\bibitem[{\citenamefont{Page et~al.}(1999)\citenamefont{Page, Brin, Motwani,
  and Winograd}}]{Page1999}
\bibinfo{author}{\bibfnamefont{L.}~\bibnamefont{Page}},
  \bibinfo{author}{\bibfnamefont{S.}~\bibnamefont{Brin}},
  \bibinfo{author}{\bibfnamefont{R.}~\bibnamefont{Motwani}}, \bibnamefont{and}
  \bibinfo{author}{\bibfnamefont{T.}~\bibnamefont{Winograd}},
  \bibinfo{type}{Technical Report}, \bibinfo{institution}{Stanford InfoLab}
  (\bibinfo{year}{1999}).

\bibitem[{\citenamefont{Toussaint}(1980)}]{Toussaint1980}
\bibinfo{author}{\bibfnamefont{G.~T.} \bibnamefont{Toussaint}},
  \bibinfo{journal}{Pattern Recognition} \textbf{\bibinfo{volume}{12}},
  \bibinfo{pages}{261} (\bibinfo{year}{1980}).

\bibitem[{\citenamefont{Gabriel and Sokal}(1969)}]{Gabriel1969}
\bibinfo{author}{\bibfnamefont{R.~K.} \bibnamefont{Gabriel}} \bibnamefont{and}
  \bibinfo{author}{\bibfnamefont{R.~R.} \bibnamefont{Sokal}},
  \bibinfo{journal}{Syst. Biol.} \textbf{\bibinfo{volume}{18}},
  \bibinfo{pages}{259} (\bibinfo{year}{1969}).

\bibitem[{\citenamefont{Sibson}(1978)}]{Sibson1978}
\bibinfo{author}{\bibfnamefont{R.}~\bibnamefont{Sibson}},
  \bibinfo{journal}{Comput. J.} \textbf{\bibinfo{volume}{21}},
  \bibinfo{pages}{243} (\bibinfo{year}{1978}).

\bibitem[{\citenamefont{Essam and Fisher}(1970)}]{Essam1970}
\bibinfo{author}{\bibfnamefont{J.~W.} \bibnamefont{Essam}} \bibnamefont{and}
  \bibinfo{author}{\bibfnamefont{M.~E.} \bibnamefont{Fisher}},
  \bibinfo{journal}{Rev. Mod. Phys.} \textbf{\bibinfo{volume}{42}},
  \bibinfo{pages}{272} (\bibinfo{year}{1970}).

\bibitem[{\citenamefont{Toroczkai and Guclu}(2007)}]{Toroczkai2007}
\bibinfo{author}{\bibfnamefont{Z.}~\bibnamefont{Toroczkai}} \bibnamefont{and}
  \bibinfo{author}{\bibfnamefont{H.}~\bibnamefont{Guclu}},
  \bibinfo{journal}{Physica A} \textbf{\bibinfo{volume}{378}},
  \bibinfo{pages}{68} (\bibinfo{year}{2007}).

\bibitem[{\citenamefont{Bertin et~al.}(2002)\citenamefont{Bertin, Billiot, and
  Drouilhet}}]{Bertin2002}
\bibinfo{author}{\bibfnamefont{E.}~\bibnamefont{Bertin}},
  \bibinfo{author}{\bibfnamefont{J.-M.} \bibnamefont{Billiot}},
  \bibnamefont{and}
  \bibinfo{author}{\bibfnamefont{R.}~\bibnamefont{Drouilhet}},
  \bibinfo{journal}{Adv. Appl. Probab.} \textbf{\bibinfo{volume}{34}},
  \bibinfo{pages}{689} (\bibinfo{year}{2002}).

\bibitem[{\citenamefont{Becker and Ziff}(2009)}]{Becker2009}
\bibinfo{author}{\bibfnamefont{A.~M.} \bibnamefont{Becker}} \bibnamefont{and}
  \bibinfo{author}{\bibfnamefont{R.~M.} \bibnamefont{Ziff}},
  \bibinfo{journal}{Phys. Rev. E} \textbf{\bibinfo{volume}{80}},
  \bibinfo{pages}{041101} (\bibinfo{year}{2009}).

\bibitem[{\citenamefont{Billiot et~al.}(2010)\citenamefont{Billiot, Corset, and
  Fontenas}}]{Billiot2010}
\bibinfo{author}{\bibfnamefont{J.~M.} \bibnamefont{Billiot}},
  \bibinfo{author}{\bibfnamefont{F.}~\bibnamefont{Corset}}, \bibnamefont{and}
  \bibinfo{author}{\bibfnamefont{E.}~\bibnamefont{Fontenas}}
  (\bibinfo{year}{2010}), \bibinfo{note}{preprint: arXiv:1004.5292}.

\bibitem[{\citenamefont{Melchert}(2013)}]{melchert2013}
\bibinfo{author}{\bibfnamefont{O.}~\bibnamefont{Melchert}},
  \bibinfo{journal}{Phys. Rev. E} \textbf{\bibinfo{volume}{87}},
  \bibinfo{pages}{042106} (\bibinfo{year}{2013}).

\bibitem[{\citenamefont{Jennings and Okino}(2002)}]{Jennings2002}
\bibinfo{author}{\bibfnamefont{E.}~\bibnamefont{Jennings}} \bibnamefont{and}
  \bibinfo{author}{\bibfnamefont{C.~M.} \bibnamefont{Okino}}, in
  \emph{\bibinfo{booktitle}{Interntional Symposium on Performance Evaluation of
  Computer and Telecommunications Systems}} (\bibinfo{year}{2002}).

\bibitem[{\citenamefont{Santi}(2005)}]{Santi2005}
\bibinfo{author}{\bibfnamefont{P.}~\bibnamefont{Santi}}, \bibinfo{journal}{ACM
  Comput. Surv.} \textbf{\bibinfo{volume}{37}}, \bibinfo{pages}{164}
  (\bibinfo{year}{2005}).

\bibitem[{\citenamefont{Li et~al.}(2005)\citenamefont{Li, Song, and
  Wang}}]{Li2005}
\bibinfo{author}{\bibfnamefont{X.-Y.} \bibnamefont{Li}},
  \bibinfo{author}{\bibfnamefont{W.-Z.} \bibnamefont{Song}}, \bibnamefont{and}
  \bibinfo{author}{\bibfnamefont{Y.}~\bibnamefont{Wang}},
  \bibinfo{journal}{Wirel. Netw.} \textbf{\bibinfo{volume}{11}},
  \bibinfo{pages}{255} (\bibinfo{year}{2005}).

\bibitem[{\citenamefont{Rajaraman}(2002)}]{Rajaraman2002}
\bibinfo{author}{\bibfnamefont{R.}~\bibnamefont{Rajaraman}},
  \bibinfo{journal}{ACM SIGACT News} \textbf{\bibinfo{volume}{33}},
  \bibinfo{pages}{60} (\bibinfo{year}{2002}).

\bibitem[{\citenamefont{Karp and Kung}(2000)}]{Karp2000}
\bibinfo{author}{\bibfnamefont{B.}~\bibnamefont{Karp}} \bibnamefont{and}
  \bibinfo{author}{\bibfnamefont{H.~T.} \bibnamefont{Kung}}, in
  \emph{\bibinfo{booktitle}{Proceedings of the 6th annual international
  conference on Mobile computing and networking}} (\bibinfo{publisher}{ACM},
  \bibinfo{year}{2000}), p. \bibinfo{pages}{243}.

\bibitem[{\citenamefont{Bose et~al.}(2001)\citenamefont{Bose, Morin,
  Stojmenovi\'{c}, and Urrutia}}]{Bose2001}
\bibinfo{author}{\bibfnamefont{P.}~\bibnamefont{Bose}},
  \bibinfo{author}{\bibfnamefont{P.}~\bibnamefont{Morin}},
  \bibinfo{author}{\bibfnamefont{I.}~\bibnamefont{Stojmenovi\'{c}}},
  \bibnamefont{and} \bibinfo{author}{\bibfnamefont{J.}~\bibnamefont{Urrutia}},
  \bibinfo{journal}{Wirel. Netw.} \textbf{\bibinfo{volume}{7}},
  \bibinfo{pages}{609} (\bibinfo{year}{2001}).

\bibitem[{\citenamefont{Yi et~al.}(2010)\citenamefont{Yi, Wan, Wang, and
  Su}}]{Yi2010}
\bibinfo{author}{\bibfnamefont{C.-W.} \bibnamefont{Yi}},
  \bibinfo{author}{\bibfnamefont{P.-J.} \bibnamefont{Wan}},
  \bibinfo{author}{\bibfnamefont{L.}~\bibnamefont{Wang}}, \bibnamefont{and}
  \bibinfo{author}{\bibfnamefont{C.-M.} \bibnamefont{Su}},
  \bibinfo{journal}{IEEE Trans. Wirel. Commun.} \textbf{\bibinfo{volume}{9}},
  \bibinfo{pages}{614} (\bibinfo{year}{2010}).

\bibitem[{\citenamefont{Kuhn et~al.}(2003)\citenamefont{Kuhn, Wattenhofer, and
  Zollinger}}]{Kuhn2003}
\bibinfo{author}{\bibfnamefont{F.}~\bibnamefont{Kuhn}},
  \bibinfo{author}{\bibfnamefont{R.}~\bibnamefont{Wattenhofer}},
  \bibnamefont{and}
  \bibinfo{author}{\bibfnamefont{A.}~\bibnamefont{Zollinger}}, in
  \emph{\bibinfo{booktitle}{Proceedings of the 2003 joint workshop on
  Foundations of mobile computing}} (\bibinfo{publisher}{ACM},
  \bibinfo{year}{2003}), p.~\bibinfo{pages}{69}.

\bibitem[{\citenamefont{Jaromczyk and Toussaint}(1992)}]{Jaromczyk1992}
\bibinfo{author}{\bibfnamefont{J.~W.} \bibnamefont{Jaromczyk}}
  \bibnamefont{and} \bibinfo{author}{\bibfnamefont{G.~T.}
  \bibnamefont{Toussaint}}, \bibinfo{journal}{Proc. IEEE}
  \textbf{\bibinfo{volume}{80}}, \bibinfo{pages}{1502} (\bibinfo{year}{1992}).

\bibitem[{\citenamefont{Barber}(1995)}]{qhull}
\bibinfo{author}{\bibfnamefont{C.~B.} \bibnamefont{Barber}},
  \emph{\bibinfo{title}{Qhull computes the convex hull, Delaunay triangulation,
  Voronoi diagram, halfspace intersection about a point, furthest-site Delaunay
  triangulation, and furthest-site Voronoi diagram.}} (\bibinfo{year}{1995}),
  \urlprefix\url{www.qhull.org}.

\bibitem[{\citenamefont{Preparata and Shamos}(1985)}]{CompGeom1985}
\bibinfo{author}{\bibfnamefont{F.~P.} \bibnamefont{Preparata}}
  \bibnamefont{and} \bibinfo{author}{\bibfnamefont{M.~I.}
  \bibnamefont{Shamos}}, \emph{\bibinfo{title}{Computational Geometry: An
  Introduction}} (\bibinfo{publisher}{Springer-Verlag}, \bibinfo{year}{1985}).

\bibitem[{\citenamefont{Kirkpatrick and Radke}(1985)}]{Kirkpatrick1985}
\bibinfo{author}{\bibfnamefont{D.~G.} \bibnamefont{Kirkpatrick}}
  \bibnamefont{and} \bibinfo{author}{\bibfnamefont{J.~D.} \bibnamefont{Radke}},
  in \emph{\bibinfo{booktitle}{Computational Geometry}}, edited by
  \bibinfo{editor}{\bibfnamefont{G.}~\bibnamefont{Toussaint}}
  (\bibinfo{publisher}{Elsevier North Holland}, \bibinfo{address}{New York},
  \bibinfo{year}{1985}), p. \bibinfo{pages}{217}.

\bibitem[{\citenamefont{Cormen et~al.}(2001)\citenamefont{Cormen, Stein,
  Rivest, and Leiserson}}]{Cormen2001}
\bibinfo{author}{\bibfnamefont{T.~H.} \bibnamefont{Cormen}},
  \bibinfo{author}{\bibfnamefont{C.}~\bibnamefont{Stein}},
  \bibinfo{author}{\bibfnamefont{R.~L.} \bibnamefont{Rivest}},
  \bibnamefont{and} \bibinfo{author}{\bibfnamefont{C.~E.}
  \bibnamefont{Leiserson}}, \emph{\bibinfo{title}{Introduction to Algorithms}}
  (\bibinfo{publisher}{MIT Press}, \bibinfo{year}{2001}).

\bibitem[{\citenamefont{Binder and Heermann}(2010)}]{Binder2002}
\bibinfo{author}{\bibfnamefont{K.}~\bibnamefont{Binder}} \bibnamefont{and}
  \bibinfo{author}{\bibfnamefont{D.~W.} \bibnamefont{Heermann}},
  \emph{\bibinfo{title}{Monte Carlo simulation in statistical physics: an
  introduction}} (\bibinfo{publisher}{Springer Science \& Business Media},
  \bibinfo{year}{2010}).

\bibitem[{\citenamefont{Houdayer and Hartmann}(2004)}]{houdayer2004}
\bibinfo{author}{\bibfnamefont{J.}~\bibnamefont{Houdayer}} \bibnamefont{and}
  \bibinfo{author}{\bibfnamefont{A.~K.} \bibnamefont{Hartmann}},
  \bibinfo{journal}{Phys. Rev. B} \textbf{\bibinfo{volume}{70}},
  \bibinfo{pages}{014418} (\bibinfo{year}{2004}).

\bibitem[{\citenamefont{Melchert}(2009)}]{autoScale2009}
\bibinfo{author}{\bibfnamefont{O.}~\bibnamefont{Melchert}}
  (\bibinfo{year}{2009}), \bibinfo{note}{preprint: arXiv:0910.5403v1}.

\bibitem[{\citenamefont{Norrenbrock}(2014)}]{norrenbrock2014}
\bibinfo{author}{\bibfnamefont{C.}~\bibnamefont{Norrenbrock}}
  (\bibinfo{year}{2014}), \bibinfo{note}{preprint: arXiv:1406.0663}.

\bibitem[{\citenamefont{Bao et~al.}(2009)\citenamefont{Bao, Cao, Wang, and
  Ding}}]{Bao2009}
\bibinfo{author}{\bibfnamefont{Z.~J.} \bibnamefont{Bao}},
  \bibinfo{author}{\bibfnamefont{Y.~J.} \bibnamefont{Cao}},
  \bibinfo{author}{\bibfnamefont{G.~Z.} \bibnamefont{Wang}}, \bibnamefont{and}
  \bibinfo{author}{\bibfnamefont{L.~J.} \bibnamefont{Ding}},
  \bibinfo{journal}{Phys. Lett. A} \textbf{\bibinfo{volume}{373}},
  \bibinfo{pages}{3032} (\bibinfo{year}{2009}).

\bibitem[{\citenamefont{Crucitti
  et~al.}(2004{\natexlab{b}})\citenamefont{Crucitti, Latora, and
  Marchiori}}]{Crucitti2004-2}
\bibinfo{author}{\bibfnamefont{P.}~\bibnamefont{Crucitti}},
  \bibinfo{author}{\bibfnamefont{V.}~\bibnamefont{Latora}}, \bibnamefont{and}
  \bibinfo{author}{\bibfnamefont{M.}~\bibnamefont{Marchiori}},
  \bibinfo{journal}{Phys. Rev. E} \textbf{\bibinfo{volume}{69}},
  \bibinfo{pages}{045104} (\bibinfo{year}{2004}{\natexlab{b}}).

\bibitem[{\citenamefont{Motter}(2004)}]{Motter2004}
\bibinfo{author}{\bibfnamefont{A.~E.} \bibnamefont{Motter}},
  \bibinfo{journal}{Phys. Rev. Lett.} \textbf{\bibinfo{volume}{93}},
  \bibinfo{pages}{098701} (\bibinfo{year}{2004}).

\bibitem[{\citenamefont{Motter and Lai}(2002)}]{Motter2002}
\bibinfo{author}{\bibfnamefont{A.~E.} \bibnamefont{Motter}} \bibnamefont{and}
  \bibinfo{author}{\bibfnamefont{Y.-C.} \bibnamefont{Lai}},
  \bibinfo{journal}{Phys. Rev. E} \textbf{\bibinfo{volume}{66}},
  \bibinfo{pages}{065102} (\bibinfo{year}{2002}).

\bibitem[{\citenamefont{Newman}(2001)}]{newman2001shortest-path}
\bibinfo{author}{\bibfnamefont{M.~E.~J.} \bibnamefont{Newman}},
  \bibinfo{journal}{Phys. Rev. E} \textbf{\bibinfo{volume}{64}},
  \bibinfo{pages}{016132} (\bibinfo{year}{2001}).

\bibitem[{\citenamefont{Ouyang et~al.}(2014)\citenamefont{Ouyang, Jin, Xia, and
  Jiang}}]{ouyang2014}
\bibinfo{author}{\bibfnamefont{B.}~\bibnamefont{Ouyang}},
  \bibinfo{author}{\bibfnamefont{X.}~\bibnamefont{Jin}},
  \bibinfo{author}{\bibfnamefont{Y.}~\bibnamefont{Xia}}, \bibnamefont{and}
  \bibinfo{author}{\bibfnamefont{L.}~\bibnamefont{Jiang}},
  \bibinfo{journal}{Eur. Phys. J. B} \textbf{\bibinfo{volume}{87}},
  \bibinfo{pages}{52} (\bibinfo{year}{2014}).

\bibitem[{\citenamefont{Hartmann}(2014)}]{Hartmann2014}
\bibinfo{author}{\bibfnamefont{A.~K.} \bibnamefont{Hartmann}},
  \bibinfo{journal}{Eur. Phys. J. B} \textbf{\bibinfo{volume}{87}},
  \bibinfo{pages}{114} (\bibinfo{year}{2014}).

\bibitem[{\citenamefont{Fisher}(1961)}]{Fisher1961}
\bibinfo{author}{\bibfnamefont{M.~E.} \bibnamefont{Fisher}},
  \bibinfo{journal}{J. Math. Phys.} \textbf{\bibinfo{volume}{2}},
  \bibinfo{pages}{620} (\bibinfo{year}{1961}).

\bibitem[{\citenamefont{Kundur et~al.}(1994)\citenamefont{Kundur, Balu, and
  Lauby}}]{kundur1994}
\bibinfo{author}{\bibfnamefont{P.}~\bibnamefont{Kundur}},
  \bibinfo{author}{\bibfnamefont{N.~J.} \bibnamefont{Balu}}, \bibnamefont{and}
  \bibinfo{author}{\bibfnamefont{M.~G.} \bibnamefont{Lauby}},
  \emph{\bibinfo{title}{Power System Stability and Control}}
  (\bibinfo{publisher}{McGraw-Hill Education}, \bibinfo{year}{1994}).

\bibitem[{\citenamefont{Filatrella et~al.}(2008)\citenamefont{Filatrella,
  Nielsen, and Pedersen}}]{filatrella2008}
\bibinfo{author}{\bibfnamefont{G.}~\bibnamefont{Filatrella}},
  \bibinfo{author}{\bibfnamefont{A.~H.} \bibnamefont{Nielsen}},
  \bibnamefont{and} \bibinfo{author}{\bibfnamefont{N.~F.}
  \bibnamefont{Pedersen}}, \bibinfo{journal}{Eur. Phys. J. B}
  \textbf{\bibinfo{volume}{61}}, \bibinfo{pages}{485} (\bibinfo{year}{2008}).

\bibitem[{\citenamefont{Rohden et~al.}(2012)\citenamefont{Rohden, Sorge, Timme,
  and Witthaut}}]{rohden2012}
\bibinfo{author}{\bibfnamefont{M.}~\bibnamefont{Rohden}},
  \bibinfo{author}{\bibfnamefont{A.}~\bibnamefont{Sorge}},
  \bibinfo{author}{\bibfnamefont{M.}~\bibnamefont{Timme}}, \bibnamefont{and}
  \bibinfo{author}{\bibfnamefont{D.}~\bibnamefont{Witthaut}},
  \bibinfo{journal}{Phys. Rev. Lett.} \textbf{\bibinfo{volume}{109}}
  (\bibinfo{year}{2012}).

\bibitem[{\citenamefont{Witthaut and Timme}(20)}]{witthaut2012}
\bibinfo{author}{\bibfnamefont{D.}~\bibnamefont{Witthaut}} \bibnamefont{and}
  \bibinfo{author}{\bibfnamefont{M.}~\bibnamefont{Timme}},
  \bibinfo{journal}{New J. Phys.} \textbf{\bibinfo{volume}{14}},
  \bibinfo{pages}{083036} (\bibinfo{year}{20}).

\end{thebibliography}

\end{document}